\documentclass[%
reprint, 
superscriptaddress, 
showpacs,
amsmath, amssymb, 
aps, 
prl,
]{revtex4-1}
\usepackage{graphicx}
\usepackage{chngcntr}
\usepackage{dcolumn}
\usepackage{bm}
\usepackage{hyperref}
\hypersetup{colorlinks=true, citecolor=blue, urlcolor=blue, linkcolor=blue}
\begin{document}
	\title{Multiplexed entanglement swapping with atomic-ensemble-based quantum memories in the single excitation regime }
	\author{Minjie Wang}
	\author{Haole Jiao}
	\author{Jiajin Lu}
	\author{Wenxin Fan}
    \author{Shujing Li} 
    \email{lishujing@sxu.edu.cn}
	\author{Hai Wang}
	\email{wanghai@sxu.edu.cn}

	\affiliation{The State Key Laboratory of Quantum Optics and Quantum Optics Devices, Institute of Opto-Electronics, Shanxi University, Taiyuan 030006,China}%
	\affiliation{Collaborative Innovation Center of Extreme Optics,
		Shanxi University, Taiyuan 030006, China}
	\begin{abstract}
Entanglement swapping (ES) between memory repeater links is critical for establishing quantum networks via quantum repeaters. So far, ES with atomic-ensemble-based memories has not been achieved. Here, we experimentally demonstrated ES between two entangled pairs of spin-wave memories via Duan-Lukin-Cirac-Zoller scheme. With a cloud of cold atoms inserted in a cavity, we produce non-classically-correlated spin-wave-photon pairs in 12 spatial modes and then prepare two entangled pairs of spin-wave memories via a multiplexed scheme. Via single-photon Bell measurement on retrieved fields from two memories, we project the two remaining memories never entangled previously into an entangled state with the measured concurrence of ${\cal C} = 0.0124 \pm 0.003$. The successful probability of ES in our scheme is increased by three times, compared with that in non-multiplexed scheme. Our presented work shows that the generation of entanglement (${\cal C} > 0$) between the remaining memory ensembles requires the average cross-correlation function of the spin-wave-photon pairs to be $ \ge 30$.
\end{abstract}

\maketitle
\section{Introduction}
Quantum networks (QNs) \cite{1,2} enable revolutionary applications including quantum simulation \cite{3}, computation \cite{4}, global quantum communication \cite{5,6}, highly-accurate frequency comparisons \cite{7,8} and long-baseline telescopes \cite{9}. To establish global QNs, quantum repeaters (QRs) are needed \cite{10,11,12}. In QRs, the long distances are divided into a number of elementary links. Entanglement is generated in each link and then successively extended via entanglement swapping between two adjacent links.

In the past two decades, entanglement between two quantum memories has been demonstrated with atomic ensembles \cite{13,14,15,16,17,18,19,20,21,22,23,24,25}, individual atoms \cite{26,27,28,29}, ions \cite{30,31,32} and solid-state spins \cite{33,34,35}, respectively. The first ES experiment was realized by using entangled photons \cite{36} via Bell state measurement (BSM). So far, ES has been widely demonstrated with photons \cite{37,38,39,40}, CV light beams \cite{41,42,43}, and hybrid CV-DV optical system \cite{44}. Recently, two-hierarchy entanglement swapping for QR has been demonstrated with four entangled pairs of photons \cite{45}. Deterministic ES with single ions \cite{46} or superconducting qubits \cite{47} has been demonstrated via quantum-logic-gate-based BSM.  

Towards to entanglement extensions in QRs, ES has been demonstrated with NV-center-based memories that interface photons via BSM on stationary qubits \cite{48,49}. Recently, Y.-F. Pu \textit{et.al.} experimentally demonstrated ES between two hybrid entanglement states between a Stokes photon and an atomic-ensemble-based memory \cite{50}, where, the Stokes photons are projected into an entangled state by performing BSM on the two retrieved photons in a post-selected way. The key of that experiment demonstrated that the scaling rate of entanglement connection between the two repeater links can be enhanced by using long-lived memories, which was a key element in QR protocols \cite{11,12,51,52,53,54}. However, even if one holds the memory-enhanced scaling, the rates of QRs are still slow for practical \cite{12, 51, 55}. To improve the slow QR rates, QRs using multiplexed memories with atomic ensembles \cite{12,55,56,57,58,59}, single ions \cite{52,60}, atomic array \cite{61}, NV centers \cite{62,63} and quantum dots with atomic-ensemble-based memories \cite{64} have been proposed. Experimentally, the multimode quantum memories based on atomic ensembles have been demonstrated \cite{58,59,65,66,67,68,69,70,71,72,73,74,75,76,77,78,79}. Heralded entanglement between two multimode quantum memories \cite{20,21} or multiplexed quantum teleportation \cite{78} from photon to memory has been demonstrated with solid-state ensembles. 

In 2007, Laurat \textit{et.al.} performed ES between two pairs of entangled quantum memories in atomic ensembles \cite{80} via Duan-Lukin-Cirac-Zoller (DLCZ) protocol. The authors firstly produce four pairs of a spin-wave (SW) memory and a Stokes photon and then prepare two entangled pairs of the ensembles. ES operation is performed by sing-photon BSM on the retrieved fields from two SW memories. The authors confirmed the creation of coherence between the remaining two memories, but didn’t detect entanglement of the memories due to SW decoherece. So far, ES with ensemble-based quantum memories remain elusive in experiment. 

Here, we experimentally demonstrated ES between two entangled pairs of SWs in atomic ensembles. Based on DLCZ process, we produce non-classically-correlated pairs of a Stokes photon and a spin wave in 12 spatial modes from a cloud of cold atoms. The multimode spin-wave-photon pairs are set to be as four DLCZ quantum interfaces A, B1, B2, C, each of them creates the spin-wave-photon pairs in three multiplexed modes. The Stokes fields from A (C) and B1 (B2) quantum interfaces are sent to a 50\% beam splitter for single-photon BSM. We then generate two entangled pairs of the spin-wave memories, which may form two elementary entangled links. Subsequently, we convert the two SWs of the interfaces B2 and B1 into anti-Stokes light fields for single-photon BSM. Conditional on a successful BSM, the remaining memories in A and C QIs are projected into an entangled state. By theoretically analyzing and experimentally measuring, we identify that the generation of entanglement (${\cal C} > 0$) between the remaining memories A and C requires for the average cross-correlation function of the four quantum interfaces above a threshold of 30, which has not been investigated previously.

\begin{figure}[htbp]
\centering
\includegraphics[width=3.3in]{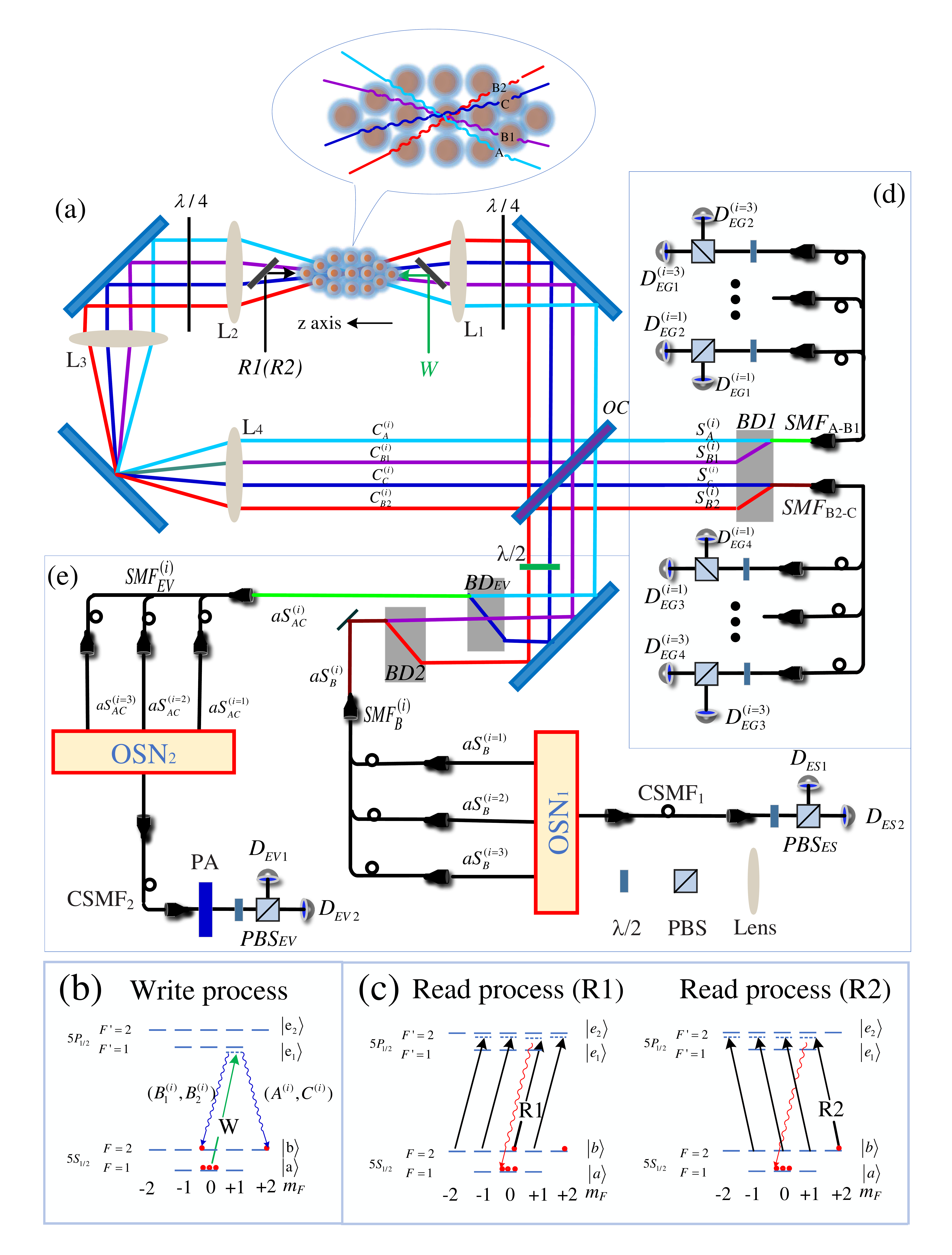}
\caption{Experimental setup. \textbf{(a)} The spatially-multiplexed source of generating the spin-wave-photon pairs. The source is formed by a cloud of Rb atoms inserted in a ring cavity. The cavity supports 12 TEM$_{00}$ modes, named as $T_A^{(i)}$, $T_{B1}^{(i)}$, $T_{B2}^{(i)}$ and $T_C^{(i)}$ cavity modes, respectively, with $i = 1,2,3$. Noted that Fig.1\textbf{(a)} is a top view of the source, meaning that we only plot four lines, labeled as $T_A^{(i)}$, $T_{B1}^{(i)}$, $T_{B2}^{(i)}$ and $T_C^{(i)}$. A bias magnetic field (100 mG) is applied along z-direction to define quantum axis. OC: optical coupler. \textbf{(b)} and \textbf{(c)} denote the relevant levels of the $^{87}$Rb atoms for the write and read processes. \textbf{(d)} The detection systems for entanglement generation (EG) of the two entangled spin-wave pairs. \textbf{(e)}: The experimental detection systems for performing ES and entanglement verification (EV). In the main text, we explained the Stokes optical circuits in \textbf{(d)} and anti-Stokes circuits in \textbf{(e)} in details. In the main text, we explained the Stokes optical circuits in d and anti-Stokes circuits in e in details. Before each of polarization beam splitters ($ PBS_{s} $), we place a half-wave plate $\left( {\lambda /2} \right)$, which is set to be ${22.5^o}$. Thus, the H- and V- polarized Stokes (anti-Stokes) fields are 50/50 mixed at the two output of PBS. In Figs \textbf{(e)} and \textbf{(d)}, the length of each fiber is about 10m.}.

\label{figure2}
\end{figure}

\section{Experimental results}
The experimental setup is shown in Fig.1a. The cloud of cold $^{87}$Rb atoms is coupled to a ring cavity, which may enhance spin-wave retrieval efficiencies \cite{81,82}. The ring cavity is a spatially-multiplexed cavity \cite{83}. It can simultaneously resonance with 12 TEM$_{00}$ modes, named as $T_A^{(i)}$, $T_{B1}^{(i)}$, $T_{B2}^{(i)}$ and $T_C^{(i)}$ modes, where, $i = 1$ to m, with $m{\rm{ = 3}}$ being the maximum storage mode number in each QI. Noted that Fig.1a is a top view, where, the modes $T_X^{(i = 1{\rm{ to 3)}}}$ are overlapped.
We carry out our ES experiment in a cyclic fashion. In the beginning of a trial, the atoms are released from a magneto-optical trap (MOT) and are prepared in the Zeeman state $\left| {a,{m_{{F_a}}} = 0} \right\rangle $. By applying a 795nm write pulse with red-detuned by 110 MHz to the $\left| a \right\rangle  \to \left| {{e_{\rm{1}}}} \right\rangle $ transition (see Fig.1b), we create spontaneous Raman emissions of Stokes photons on $\left| {{e_{\rm{1}}}} \right\rangle  \leftrightarrow \left| b \right\rangle $ transition. The write pulse is  ${\sigma ^ + }$-polarized and directs into the atoms through a beam splitter \textit{BS$_{W}$} along z axis. It induces the Stokes-photon emissions on   ${\sigma ^ + }$ -transition $\left| {{e_1},{m_F} =  + {\rm{1}}} \right\rangle  \leftrightarrow \left| {b,{m_F} = 0} \right\rangle $ or  ${\sigma ^ - }$  -transition $\left| {{e_1},{m_F} =  + {\rm{1}}} \right\rangle  \leftrightarrow \left| {b,{m_F} =  + {\rm{2}}} \right\rangle $, which simultaneously create spin waves associated with the Zeeman coherence $\left| {a,{m_{Fa}} = {\rm{0}}} \right\rangle  \to \left| {b,{m_{Fb}} = {\rm{0}}} \right\rangle $  or $\left| {a,{m_{Fa}} = {\rm{0}}} \right\rangle  \to \left| {b,{m_{Fb}} =  + {\rm{2}}} \right\rangle $. 
When a  $ {\sigma ^ - } $- ($ {\sigma ^ + } $ ) polarized Stokes photon is emitted into the $T_A^{(i)}$ or $T_C^{(i)}$ ($T_{B1}^{(i)}$ or $T_{B2}^{(i)}$  )  cavity mode and propagates along counterclockwise, it will be collected into the entanglement-generation detection system in Fig.1d (see below explains) and then is taken as Stokes photon $S_A^{(i)}$  or $S_C^{(i)}$  ($S_{B1}^{(i)}$ or$S_{B2}^{(i)}$). Along with the creation of $S_A^{(i)}$, $S_{B1}^{(i)}$, $S_{B2}^{(i)}$ or $S_C^{(i)}$ photon, one collective excitation is created and stored in the spin waves $ A_{}^{(i)} $ ,$ B1_{}^{(i)} $  ,$ B2_{}^{(i)} $  or $ C_{}^{(i)} $   with their wave-vectors defined by $ k_{A,B1,B2,C}^{(i)} = {k_w} - k_{{T_{A,B1,B2,C}}}^{(i)} $, where, ${k_w}$  is the wave-vector of the write pulse and $ k_{{T_{A,B1,B2,C}}}^{(i)} $ that of the Stokes photon propagating in the  $T_A^{(i)}$ , $T_{B1}^{(i)}$ , $T_{B2}^{(i)}$ or $T_C^{(i)}$ cavity mode. The spin waves $ A_{}^{(i)} $ and $ C_{}^{(i)} $ ($ B1_{}^{(i)} $  ,$ B2_{}^{(i)} $) are associated with the coherence $ \left| {a,{m_{Fa}} = {\rm{0}}} \right\rangle  \to \left| {b,{m_{Fb}} = {\rm{0}}} \right\rangle $   ($ \left| {a,{m_{Fa}} = {\rm{0}}} \right\rangle  \to \left| {b,{m_{Fb}} =  + {\rm{2}}} \right\rangle $  ). The probability of generating the spin-wave-photon pair in each of 12 modes is set to be $ \chi    $, which is  far less than 1 in order to avoid multi-excitations\cite{12}. At the left of the atoms, We inserted a wave-plate ($\lambda /4$) to transform ${\sigma ^ - }$ - (${\sigma ^ + }$ ) polarized Stokes photons in the modes $T_A^{(i)}$ and $T_C^{(i)}$ ($T_{B1}^{(i)}$ and $T_{B2}^{(i)}$  ) into V- (H-) polarized ones, where, $i = 1$ to $ m = 3$. As shown in Fig.1a, the Stokes photon in the $T_X^{(i)}$ modes may escape from the cavity through the optical coupler (OC) and direct into the $S_X^{(i)}$  mode. The V-polarized $S_A^{(i)}$ ($S_{B2}^{(i)}$) and H-polarized $S_{B1}^{(i)}$ ($S_{C}^{(i)}$) fields are combined at a beam displacer (BD1), which are then coupled into a single-mode fiber ${\rm{SMF}}_{A - B1}^{(i)}$ (${\rm{SMF}}_{B2 - C}^{(i)}$ ) and then sent to the detection system for entanglement generation (EG) (Fig.1d). For our multiplexed scheme, the Stokes fields in the modes ${S_X}^{(i = {1_{}}t{o_{}}m)}$  (m=3) are served as S$_X$ Stokes photon, and the spin waves $X_{}^{(i)}$, which are correlated with the Stokes fields propagating in $T_X^{(i)}$, respectively, are served as X (=A, B1, B2 or C) spin wave ($i = {\rm{1 }}$ to 3). Since the spin waves A, B1, B2, and C are distinguishably stored, they define the individual memory ensembles \cite{18,25,84,85,86}, also named as A, B1, B2, and C. Furthermore, A, B1, B2, and C memory ensembles also are regard as quantum interfaces (QIs) since they can individually produce spin-wave-photon pairs. In this way, upon the detection events simultaneously occurred at the detectors $D_{EG1}^{(i)}$ and $D_{EG3}^{(i)}$ ($i = {\rm{1 }}$ to 3 ). The two pairs of memory ensembles A-B1 and B2-C are projected into entangled states $ {\Psi _{A,B1}} = \left( {\left| 1 \right\rangle {}_A{{\left| 0 \right\rangle }_{B1}} \pm {e^{ - i\Delta {\alpha _1}}}\left| 0 \right\rangle {}_A{{\left| 1 \right\rangle }_{B1}}} \right)/\sqrt 2 $  and $ {\Psi _{B2,C}} = \left( {\left| 1 \right\rangle {}_{B2}{{\left| 0 \right\rangle }_C} \pm {e^{ - i\Delta {\alpha _{_2}}}}\left| 0 \right\rangle {}_{B2}{{\left| 1 \right\rangle }_C}} \right)/\sqrt 2  $, respectively, where, $  \Delta {\alpha _1} = {\alpha _A} - {\alpha _{B1}} $ ($\Delta {\alpha _2} = {\alpha _{B2}} - {\alpha _C}$ ) is the phase difference between Stokes fields \textit{$ S_{A} $} and \textit{$ S_{B1} $} (\textit{$ S_{B2} $} and \textit{$ S_{C} $}) in the propagations from the memory ensembles to \textit{$ PBS_{EG1} $} ({$ PBS_{EG1} $}). The use of the multiplexed scheme enable the probability of generating $ {\Psi _{A,B1}} $  (  $ {\Psi _{B2,C}} $) to reach $ 3\chi $  ($ 3\chi $), which is three times that of the non-multiplexed scheme. 

After a storage time \textit{t$ _{1} $}, we apply a ${\sigma ^ + }$ -polarized read pulse \textit{R1} to converts the spin waves B1 and B2 into anti-Stokes fields, which propagate in $T_{B1}^{(i)}$  and $T_{B2}^{(i)}$  cavity modes along clockwise \cite{85}. Escaping from the cavity mirror OC, the anti-Stokes fields in $T_{B1}^{(i)}$  and $T_{B2}^{(i)}$ modes direct into $aS_{B1}^{(i)}$  and $aS_{B2}^{(i)}$ channels, which are H-polarized and V- polarized, respectively. We also use BD2 to combine $aS_{B1}^{(i)}$  and $aS_{B2}^{(i)}$ fields into spatial modes $aS_{B1 - 2}^{(i)}$. The modes $aS_{B1 - 2}^{(i)}$ are coupled into single-mode fibers ${\rm{SMF}}_{B1 - 2}^{(i)}$. After passing through ${\rm{SMF}}_{B1 - 2}^{(i)}$, the $aS_{B1 - 2}^{(i = {1_{}}t{o_{}}3)}$ fields are sent to an optical switch network (OSN$_1$) \cite{75} for their circuits multiplexing. Based on a feed-forward signal, the OSN$_1$ routes $aS_{B1 - 2}^{(i)}$ fields into a common single-mode fiber (CSMF$_1$). After CSMF$_1$, $ a{S_{B1 - 2}}^{(i)} $  fields are sent to a polarization beam-splitter ($ PBS_{ES} $) the detection system for ES. The outputs $\left( {{a_{a{S_{B1}}}} \pm {e^{i\Delta \gamma }}{a_{a{S_{B2}}}}} \right)/\sqrt 2 $  from \textit{PBS$_{ES}$} direct into the single-photon detectors {D$_{ES1}$}  and {D$_{ES2}$}(Fig.2e), where, ${a_{a{S_{B1,2}}}}$  denote the annihilation operators associated with the fields $a{S_{B1,2}}$,  $\Delta \gamma $ the phase difference between the fields  and  from the memory ensembles to {PBS$_{ES}$}. Upon a detection event at the detector {D$_{ES1}$}\cite{80,86}, the remaining spin-wave memories A and C are projected into the state.
\begin{equation}
	{\rho _{A,C}} = \left| 0 \right\rangle \left\langle 0 \right|/2 + \left| {{\Psi _{A,C}}} \right\rangle \left\langle {{\Psi _{A,C}}} \right|/2
\end{equation}
where, $\left| 0 \right\rangle $  is vacuum part, ${\Psi _{A,C}} = \left( {\left| 1 \right\rangle {}_A{{\left| 0 \right\rangle }_C} \pm {e^{ - i\xi }}\left| 0 \right\rangle {}_A{{\left| 1 \right\rangle }_C}} \right)/\sqrt 2 $ represents entangled state between A and C memory ensembles, $\xi  = \Delta {\alpha _1} - \Delta {\alpha _2} + \Delta \gamma  $  sum of the difference phases, which is passively stabilized and kept constant in our presented experiment. The entanglement state $ {\Psi _{A,C}} $  can be characterized by the concurrence \cite{13,23}:     
\begin{equation}
	\begin{array}{l}
		{\cal C} = \max \{ 0,\left( {\left( {{p_{10}} + {p_{01}}} \right)V - 2\sqrt {{p_{00}}{p_{11}}} } \right)/P\\
		{\rm{  }} \approx \max \left\{ {0,{p_c}\left( {V - \sqrt h } \right)} \right\}
	\end{array}
\end{equation}
where, \textit{$V_{AC}  $} denotes the visibility of interference between the retrieved (anti-Stokes) fields $a{S_A}$ and $a{S_C}$, ${p_{ij}}$  corresponds to the probability to find $i \in \left( {0,1} \right)$  photons in the retrieved field $a{S_A}$, and  $j \in \left( {0,1} \right)$ photons in the retrieved field $a{S_C}$, $P = {p_{10}} + {p_{01}} + {p_{11}} + {p_{00}} \approx 1$, $h = {p_{11}}/\left( {{p_{10}}{p_{01}}} \right) < 1$  denotes the suppression parameter of two-photon events relative to the square of the probability of single-photon events \cite{13}, ${p_c} = {p_{10}} + {p_{01}} \approx 2{p_{10}} \approx 2{p_{01}}$, ${\cal C}$ takes values from 0 for a separable state to 1 for a maximally entangled state. 

To verify entanglement between the remaining memories A and C, we measure the concurrence ${\cal C}$ of the entangled state ${\Psi _{A,C}}$. For which, we apply a  ${\sigma ^ - }$-polarized read pulse \textit{R2} at the storage time  \textit{t$ _{2} $} (${t_2} = {t_1} + \Delta t$).
The \textit{R2}  convert the A and C spin waves into anti-Stokes fields, which are H- and V- polarized and propagate in   $T_A^{(i)}$ and $T_C^{(i)}$ modes, respectively, along clockwise. The anti-Stokes fields in $T_A^{(i)}$ and $T_C^{(i)}$ cavity modes escape from the cavity OC, which then direct into $aS_{A}^{(i)}$ and $aS_{C}^{(i)}$ channels, respectively. The H- and V- polarized fields $aS_{A}^{(i)}$ and $aS_{C}^{(i)}$ are combined into the spatial mode $aS_{AC}^{(i)}$ after the beam displacer (BD$_{EV}$). The fields in $aS_{AC}^{(i)}$ mode are coupled into the single-mode fiber ${\rm{SMF}}_{EV}^{(i)}$ and then are sent to another optical switch network OSN$_2$. OSN$_2$ routed the $aS_{AC}^{(i)}$ modes ($i = 1{\rm{ to 3}}$) into a common single-mode fiber CSMF$_2$ \cite{75}. After the CSMF$_2$, the H-polarized and V-polarized fields both in the spatial mode $aS_{AC}^{}$ go through a phase adjustment (PA), which can vary the relative phase $\theta $ between the two fields \cite{25}. The fields in  $aS_{AC}^{}$  then are sent to the detection system, which includes a polarization beam-splitter (\emph{PBS}$_{EV}$) and two single-photon detectors  ${D_{EV1}}$ and ${D_{EV2}}$(Fig.1e). The visibility {$ V_{AC}  $} can be measured by detecting the coincidence counts between the detectors  ${D_{ES1}}$ and  ${D_{EV1}}$ \cite{86} with  $\theta $  being scanned. While, the suppression parameter \textit{$ h_{AC}  $} can be measured by directly detecting the photon counts in the \textit{$ aS_{A} $} and \textit{$ aS_{C} $} fields.
 \begin{figure}[h]
	\centering\includegraphics[width=8cm]{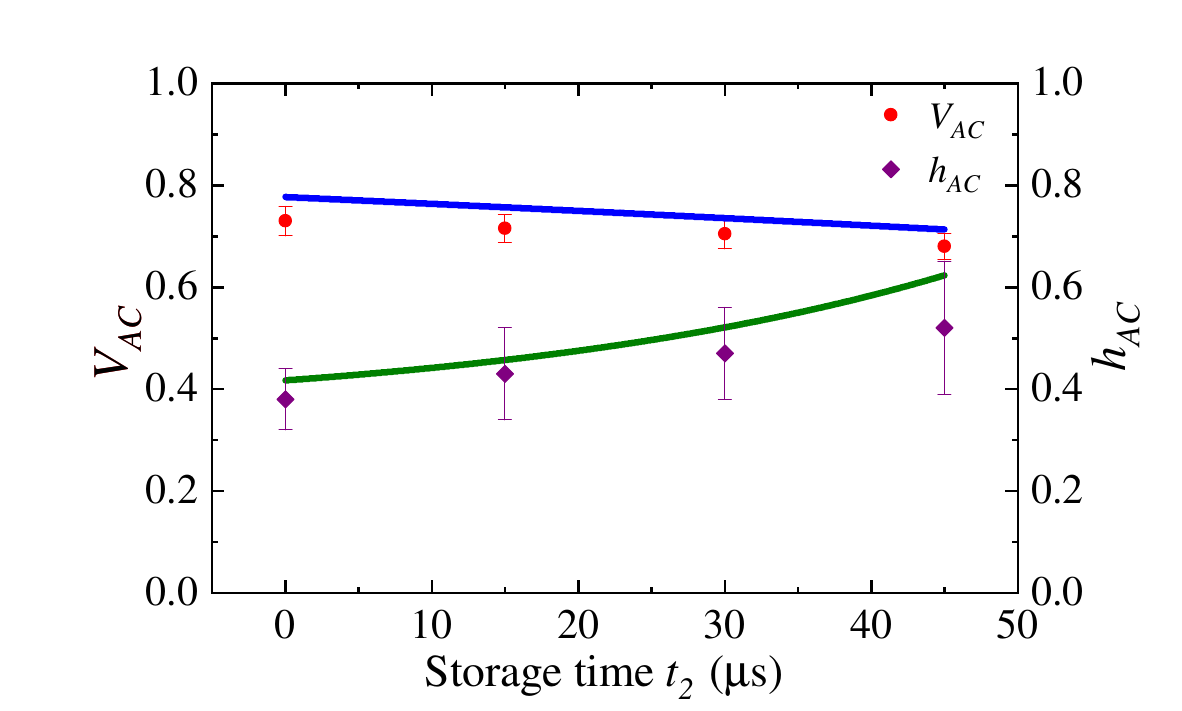}
	\caption{The measured \textit{V$_{AC}$} (red circles) and \textit{h$_{AC}$} (purple diamonds) as a function of the storage time \textit{t$ _{2} $}, where,${t_2} = {t_1} + 2\mu s$, which is changed by varying the time \textit{t$ _{1} $}, at which we perform ES.}
\end{figure}
In previous works, the quantitative relationship between the visibility\textit{$ V_{AC} $} of the entangled state $ {\Psi _{A,C}} $ and the cross-correlation functions of the spin-wave-photon pairs has not been pointed out. In the presented work, we make the detailed theoretical analyses on the relationship in Supplementary material \cite{86} and write the main results as:
 \begin{equation}
	{V_{AC}} \approx 1 - 4/{g_{S,aS}}({t_1}) - 4/g{'_{S,aS}}({t_2})
\end{equation}
\begin{equation}
	{h_{AC}} \approx 8 \times \left( {1/{g_{S,aS}}({t_1}) + 1/g{'_{S,aS}}({t_2})} \right) 
\end{equation}
where, $g_{S,aS}{({t_1})}$ ($g_{S,aS}'{({t_2})}$) denotes the cross-correlation function for either one spin-wave-photon pairs from B1 (A) and B2 (C) memory ensembles at time ${t_1}$(${t_2}$), with $g_{S,aS}^{(B1)}({t_1}) = g_{S,aS}^{(B2)}({t_1}) = g_{S,aS}^{}({t_1})$ ($g_{S,aS}^{(A)}({t_2}) = g_{S,aS}^{(C)}({t_2}) = {g_{S,aS}}({t_2})$) assumed. 
\begin{figure}[h]
	\centering\includegraphics[width=8cm]{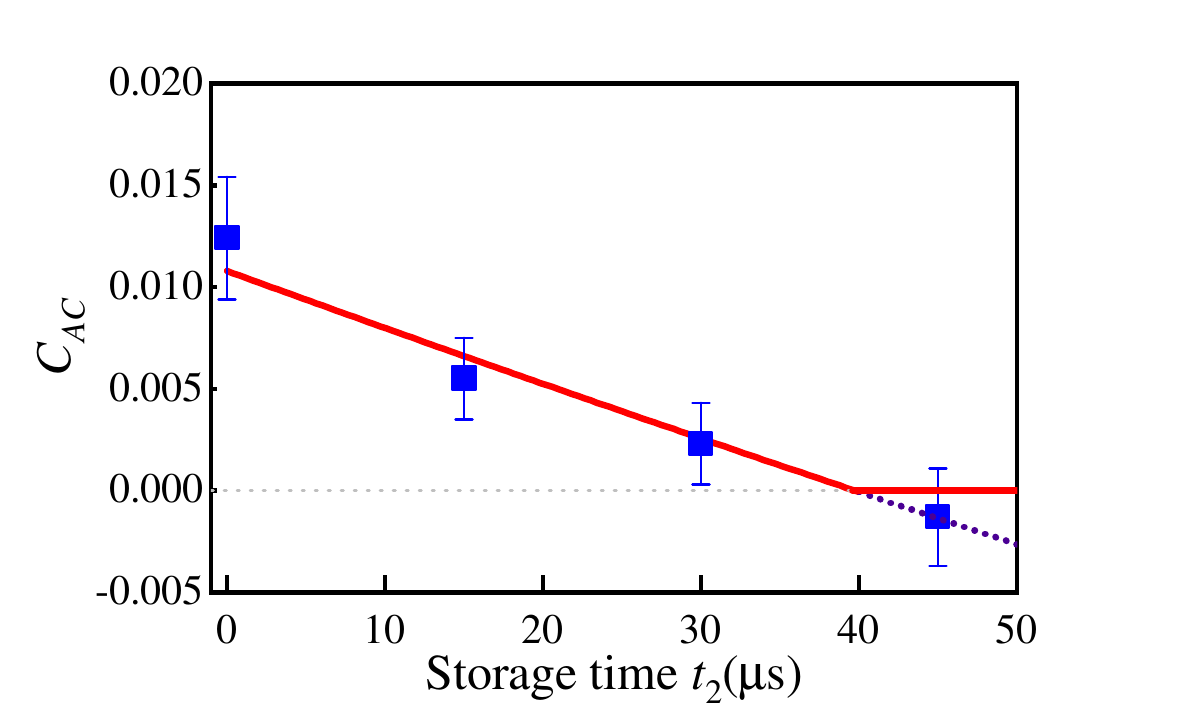}
	\caption{The measured concurrence ${{\cal C}_{AC}}$ (blue squares) as a function of the storage time \textit{t$ _{2} $}.}
\end{figure}
With Eq. (2) as well as Eqs.(3), (4), we calculated the threshold of the cross-correlations $g_{S,aS}^{}$ to generate entanglement (the concurrence ${\cal C} \ge 0$) between the two remaining memories, which is $g_{S,aS}^{} \approx 29.3$.

    The measured \textit{V$_{AC}$ }(red circles) and \textit{h$_{AC}$} (purple diamonds) data as a function of the storage time \textit{t$ _{2} $} are shown in Fig.2, respectively. The solid curves in Fig. 2 are the fits to the measured V$_{AC}$ and \textit{h$_{AC}$} data according to ${V_{AC}} \approx 1 - 4/{g_{S,aS}}({t_1}) - 4/g{'_{S,aS}}({t_2})$ and ${h_{AC}} = 8 \times \left( {1/{g_{S,aS}}({t_1}) + 1/g{'_{S,aS}}({t_2})} \right)$, respectively. In these fittings, the data of ${g_{S,aS}}({t_1})$, $g{'_{S,aS}}({t_2})$ are taken from the fittings to the measured ${g_{S,aS}}({t_1})$, $g{'_{S,aS}}({t_2})$ data, which are presented in Fig. 2S in Supplemental material \cite{86}. One can see that the fittings in Fig. 2 are in agreement with the measured \textit{V$_{AC}$ } and \textit{h$_{AC}$} data.

    Based on the measured \textit{V$_{AC}$ } and \textit{h$_{AC}$} data as well as ${p_c} = {p_{10}} + {p_{01}}$ for \textit{aS$_{AC}$} fields, we presented the concurrence ${{\cal C}_{AC}}$  (blue squares) as a function of the storage time \textit{V$_{AC}$ } according to Eq.(2) in Fig. 3. The solid red curve is the fit to the measured ${{\cal C}_{AC}}$  data according to the relationship
    \begin{widetext}
    \centering
    ${{\cal C}_{AC}} \approx \max \left\{ {0,{\rm{ }}{p_c}\left( {1 - 4/{g_{S,aS}}({t_1}) - 4/g{'_{S,aS}}({t_2}) - \sqrt {8 \times \left( {1/{g_{S,aS}}({t_1}) + 1/g{'_{S,aS}}({t_2})} \right)} } \right)} \right\}$ 
    \end{widetext}
    From Fig. 3, one can see that the concurrence ${{\cal C}_{AC}}$  decreases with \textit{t$ _{2} $}, the reason for this is decoherence of the spin waves. At ${t_2} = 32\mu s$, the concurrence ($0.0023 \pm 0.002$) is still beyond zero. On the other hand, the average of ${g_{S,aS}}({t_1})$ and $g{'_{S,aS}}({t_2})$ is $ \sim $30 at this time \cite{86}, which is basically in agreement with the theoretically-predicted cross-correlation function threshold of the $g_{S,aS}^{} \approx 29.3$.

    The Fig.4 presents the measured four-photon coincidence counts between the detectors ${D_{EG1}}$,${D_{EG3}}$,${D_{ES1}}$,${D_{EV1}}$  as a function of the number of the storage modes \textit{m}, which present a linear increase with the increase in \textit{m}, showing the multiplexing potential in ES operation. 
    \begin{figure}[ht]
    	\centering\includegraphics[width=7cm]{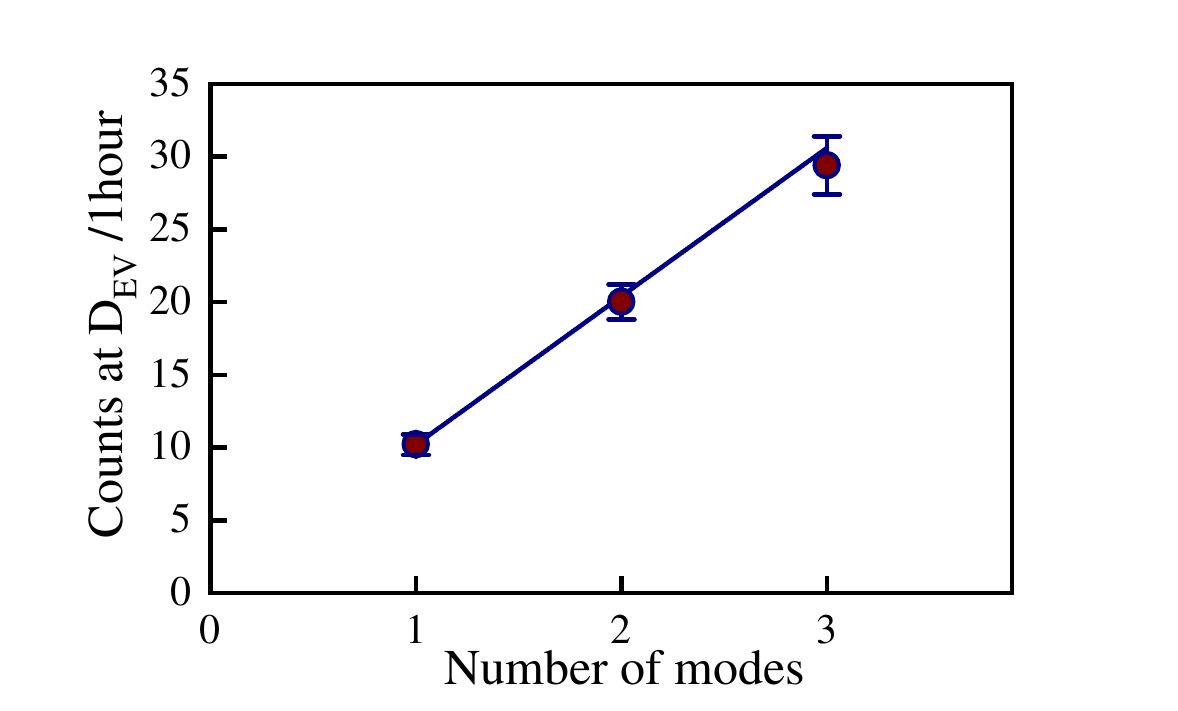}
    	\caption{The measured four-photon coincidence counts (blue circles) between the detectors ${D_{EG1}}$,${D_{EG3}}$,${D_{ES1}}$,${D_{EV1}}$ as a function of the number m of the storage modes.}
   \end{figure}

\section{Conclusion}
We experimentally demonstrated ES between two entangled pairs of memory ensembles via a spatially-multiplexed DLCZ protocol. At ${\rm{2-\mu s}}$ delay, the measured concurrence of the entanglement state between the remaining memory ensembles A and C after ES is ${{\cal C}_{AC}} = 0.124 \pm 0.03$ , which violates the occurrence inequality ${\cal C} \ge 0$  by 4.1 standard deviations and then represent successful ES. The use of the multi-mode memories in our scheme increase ES rate by a factor of 3, compared with that of the scheme using the non-multiplexed memories. 
We theoretically and experimentally presented the first study on the quantitative connection between the average cross-correlation function ${\bar g_{S,aS}}$  of the spin-wave-photon pairs and the concurrence ${\cal C}$  of the entanglement state between the remaining memory ensembles. The measured data are well in agreement with the theoretical results as expected. Both indicate that the threshold of the concurrence   occurs ${{\cal C}} = 0$ at ${\bar g_{S,aS}} \approx 30$, which is 4-times the threshold $ {\bar g_{S,aS}} \approx 7 $ required for generating the single entangled links \cite{84,87}, showing that the resultant entanglement via ES requires QIs to have more high performances. A maximum storage time, known as cutoff, is enforced on the memories to ensure to generate high-quality entanglement \cite{88,89,90,91,92} after ES. Our study shows that the cutoffs for ES with ensembles are mainly determined by the dependence of $ {\bar g_{S,aS}}  $  on the storage time. We noted that single repeater links distributed single-excitation entanglement between two atomic ensembles separated by metropolitan fibers have been demonstrated recently \cite{22}. In the future, ES between two or more repeater links over metropolitan fibers are now envisioned \cite{5, 93}. Our research result provides an important reference for setting the policy in practically realizing the metropolitan-fiber repeater links. Aiming at the goal, memories must have high-performance: including long-lived lifetimes \cite{81}, low-background noise, massively- multiplexed storages \cite{72}, and quantum-frequency conversation \cite{19,22,28,94}. Beside using high-performance memories, entanglement purification is required in the ES. With the above improvements, entanglement connection between two or more repeater links over a significant distance will be experimentally realized.

\part*{{\normalsize {\Large Acknowledgements}}}

This work is supported by Key Project of the Ministry of Science and Technology of China (Grant No. 2016YFA0301402), the National Natural Science Foundation of China (12174235), the Fund for Shanxi Key Subjects Construction (1331), and the Fundamental Research Program of Shanxi Province (202203021221011).

\begin{widetext}
\clearpage
\title{Supplementary material }
\section{I. The measured intrinsic retrieval efficiencies as functions of the storage time t for the four memory ensembles}
The measured intrinsic retrieval efficiencies as functions of the storage time $t$  for spin waves A (black squares), B1 (blue triangles), B2 (red circles) and C (green diamonds) are presented in Fig.S1, which are shown that retrieval efficiencies of the four memories are approximately identical, i.e., ${\gamma ^{(A)}} \approx {\gamma ^{(B1)}} \approx {\gamma ^{(B2)}} \approx {\gamma ^{(C)}} \approx \gamma $. The solid curve is the fitting to the measured results according to $\gamma (t) = {\gamma _0}{e^{ - t/{\tau _0}}}$, yielding ${\gamma _0} = 68\% $, ${\tau _0} = 320\mu s$.

\begin{figure}[ht]
	\centering\includegraphics[width=9cm]{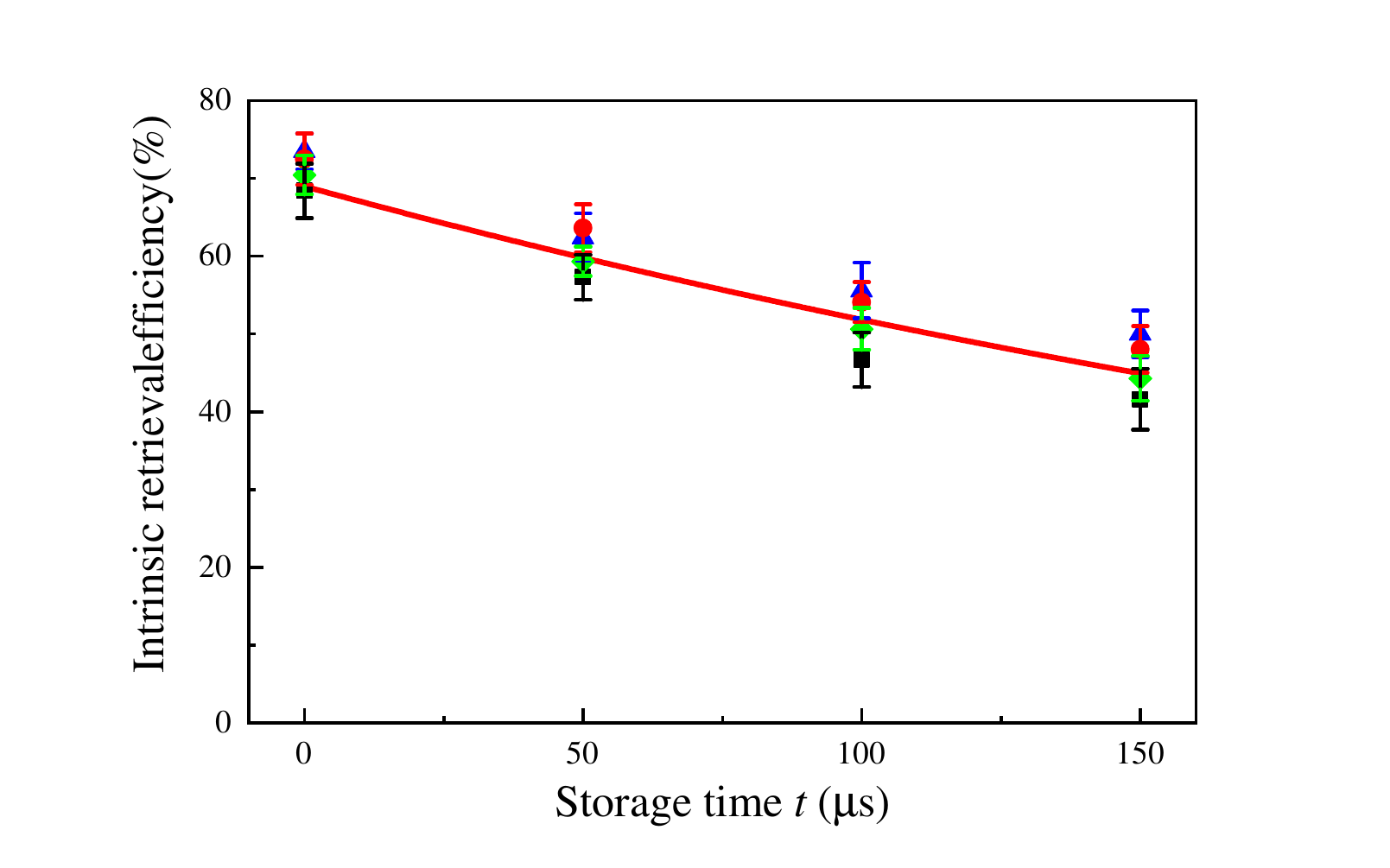}

	\textbf{Fig. 1S.} The measured intrinsic retrieval efficiencies as a function of the storage time t for spin-wave modes A, B1, B2 and C. The red solid curve is the fitting to the measured results according to  $\gamma (t) = {\gamma _0}{e^{ - t/{\tau _0}}}$, yielding ${\gamma _0} = 68\% $, ${\tau _0} = 320\mu s$.
	\label{S1}
\end{figure}
\section{II. Cross-correlation functions of the four quantum interfaces A, B1, B2, C }
The cross-correlation function of X quantum interface (X= A, B1, B2, C) is defined by $g_{S,aS}^{(X)} = P_{S,aS}^{(X)}/(P_S^{(X)}P_{aS}^{(X)})$, where, $P_S^{(X)}$ ($P_{aS}^{(X)}$) denotes the probability of detecting a Stokes (anti-Stokes) photon in $ S_{X} $ (in $ aS_{X} $) mode, $P_{S,aS}^{(X)}$  the coincident probability between the Stokes field $ S_{X} $ and the retrieved (anti-Stokes) field $ aS_{X} $.

We now present the dependence of the cross-correlation functions on the retrieval efficiency and the background noise in the detection channels. $P_S^{(X)}$  and $P_{aS}^{(X)}$, which represent unconditional probabilities of detecting a photon in$  S_{X}  $ or $ aS_{X} $ mode, can be written as [1] 
\setcounter{equation}{0}
\renewcommand{\theequation}{\thesection S\arabic{equation}}
\begin{subequations}
\begin{equation}
	P_S^{(X)} = \chi \eta 
\end{equation}
\begin{equation}
	P_{aS}^{(X)} = \chi \gamma \left( t \right)\eta  + \chi \left( {1 - \gamma (t)} \right){\xi _{se}}f\eta  + {Z^{(X)}}\eta 
\end{equation}
\end{subequations}
where,  $ \eta $  is the detection efficiency for either detection channel alone, with ${\eta ^{(A)}} = {\eta ^{(B1)}} = {\eta ^{(B2)}} = {\eta ^{(C)}} = \eta $, $\gamma (t)$ is the retrieval efficiency for either quantum interface alone, with ${\gamma ^{(A)}}(t) = {\gamma ^{(B1)}}(t) = {\gamma ^{(B2)}}(t) = {\gamma ^{(C)}}(t) = \gamma (t)$, $f = F/2\pi $ is the cavity enhance factor with \textit{F}  being the cavity finesse [1], ${\xi _{se}}$ the branching ratio corresponding to the read-photon transitions[2, 3], ${Z^{(X)}}$  the background noise in the anti-Stokes mode  $ aS_{X} $.  Noted that the background noise in the Stokes mode $ S_{X} $ is very small and neglected in the presented scheme. For DLCZ quantum memory, the coincident probability $P_{S,aS}^{(X)}$  can be written as[4]
\begin{equation}
	P_{S,aS}^{(X)}(t) = \left\langle {P_S^{(X)}P_{aS}^{(X)}} \right\rangle  = P_S^{(X)}\gamma (t)\eta  + P_S^{(X)}P_{aS}^{(X)} = \chi \gamma (t){\eta ^2} + P_S^{(X)}P_{aS}^{(X)}
\end{equation}
The cross-correlation function of X can be rewritten as:
\begin{equation}
g_{S,aS}^{(X)} = 1 + \frac{{\gamma (t)\eta }}{{P_{aS}^{(X)}}}
\end{equation}
Introducing the expression of $P_{aS}^{(X)}$  into the above equation, we have:
\begin{equation}
	g_{S,aS}^{(X)} = 1 + \frac{{\gamma \left( t \right)}}{{\chi \gamma \left( t \right) + {Z^{(X)}} + \chi \left( {1 - \gamma (t)} \right){\xi _{se}}f}}
\end{equation}
Based on Eq.(S3), we have:
\begin{equation}
	P_{aS}^{(X)} = \frac{{\gamma (t)\eta }}{{g_{S,aS}^{(X)} - 1}} \approx \frac{{\gamma (t)\eta }}{{g_{S,aS}^{(X)}}}
\end{equation}
In our presented experiment, we store the spin waves B1 and B2 (A and C) as the same spin transition $ \left| {g,{m_F} = 0} \right\rangle  \leftrightarrow \left| {s,{m_F} = 0} \right\rangle $   ( $\left| {g,{m_F} = 0} \right\rangle  \leftrightarrow \left| {s,{m_F} = 2} \right\rangle $ ). The used read pulse to retrieve B1 and B2 (A and C) spin waves are the same, which is $ R_{1} $ ($ R_{2} $) laser pulse, as shown in Fig.2 in the main text. In this way, the background noise in the anti-Stokes $ aS_{B1} $ and $ aS_{B2} $ ($ aS_{A} $ and $ aS_{C} $) modes are the approximately same, i.e., ${Z^{(B1)}} \approx {Z^{(B2)}} = Z$  (${Z^{(A)}} \approx {Z^{(C)}} = Z'$).The background noise in anti-Stokes $ aS_{A} $ ($ aS_{C} $) mode is different from that in $ aS_{B1} $ ($ aS_{B2} $) mode (channel). The reason for this is that we apply R1 pulse to retrieve the spin waves B1 and B2 before the retrievals of the spin waves A and C, thus, the application of $ R_{1} $ pulse introduces an additional noise into A and C spin waves. We then have  
\begin{subequations}
	\begin{equation}
	P_{aS}^{(B1)} = P_{aS}^{(B2)} = P_{aS}^{} = \chi \gamma \left( t \right)\eta  + \chi \left( {1 - \gamma (t)} \right){\xi _{se}}f\eta  + Z\eta 
	\end{equation}
	\begin{equation}
	P_{aS}^{(A)} = P_{aS}^{(C)} = P_{aS}^{'} = \chi \gamma \left( t \right)\eta  + \chi \left( {1 - \gamma (t)} \right){\xi _{se}}f\eta  + Z'\eta 
	\end{equation}
\end{subequations}
The measured cross-correlation functions for A, B1, B2 and C quantum interfaces are presented in Fig.2S, where, the black squares, blue triangles, red circles and green diamonds are the data for A, B1, B2 and C spin-wave modes, respectively. One can see that the measured data for B1 and B2 (A and C) modes are approximately identical, which is well in agreement with the above analysis. The curve F1 is the fitting to the measured data for B1 and B2 quantum interfaces according to
\begin{subequations}
	\begin{equation}
		g_{S,aS}^{(B1)}(t) = g_{S,aS}^{(B2)}(t) = g_{S,aS}^{}(t) = 1 + \frac{{\gamma \left( t \right)}}{{\chi \gamma \left( t \right) + Z + \chi \left( {1 - \gamma (t)} \right){\xi _{se}}f}} 
	\end{equation}
While, the curve F2 is that for A and C quantum interfaces according to 
	\begin{equation}
		g_{S,aS}^{(A)}(t) = g_{S,aS}^{(C)}(t) = g'_{S,aS}(t) = 1 + \frac{{\gamma \left( t \right)}}{{\chi \gamma \left( t \right) + Z' + \chi \left( {1 - \gamma (t)} \right){\xi _{se}}f}}
	\end{equation}
\end{subequations}
where, $Z \approx 1*{10^{ - 3}}$  and $Z' \approx 3*{10^{ - 3}}$  are taken from experimentally measured data. The other parameters see the caption in Fig.S2.
\begin{figure}[ht]
	\centering\includegraphics[width=9cm]{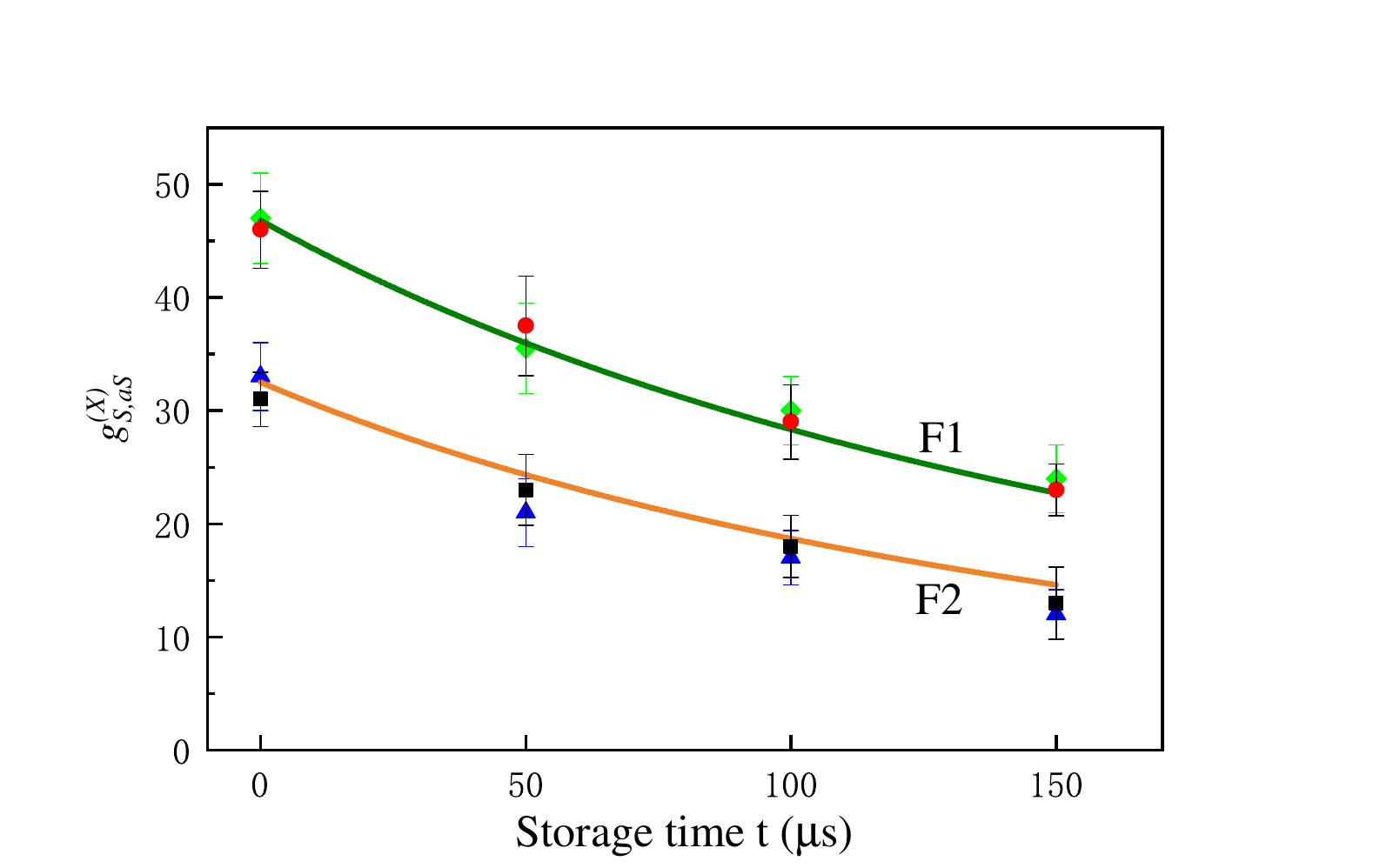}

	\textbf{Fig. 2S.} The measured cross-correlation functions as function of the storage time t for A (black squares), C (blue triangles), B1 (red circles) and B2 (green diamonds) quantum interfaces. Solid curve F1 (F2) is the fitting to the measured data for B1 and B2 (A and C) quantum interfaces according to the Eq. (S7a, b), where the values of parameters $\chi  = 0.01$ ,${Z^{(B1)}} = {Z^{(B2)}} = Z \approx 1*{10^{ - 3}}$, ${\zeta _{se}} = 0.3$, $f = 10$  ($\chi  = 0.01$, ${Z^{(A)}} = {Z^{(C)}} = Z' \approx 3*{10^{ - 3}}$, ${\zeta _{se}} = 0.3$, $f = 10$) are taken from experimentally measured data. The retrieval efficiency  $\gamma \left( t \right)$ is taken from the fitting to the measured results in Fig.1S.
	\label{S2}
\end{figure}
\section{III.Dependence of the visibility V of the entangled state   on the cross-correlation functions of the spin-wave-photon pairs from the four quantum interfaces}
\begin{figure}[ht]
	\centering\includegraphics[width=9cm]{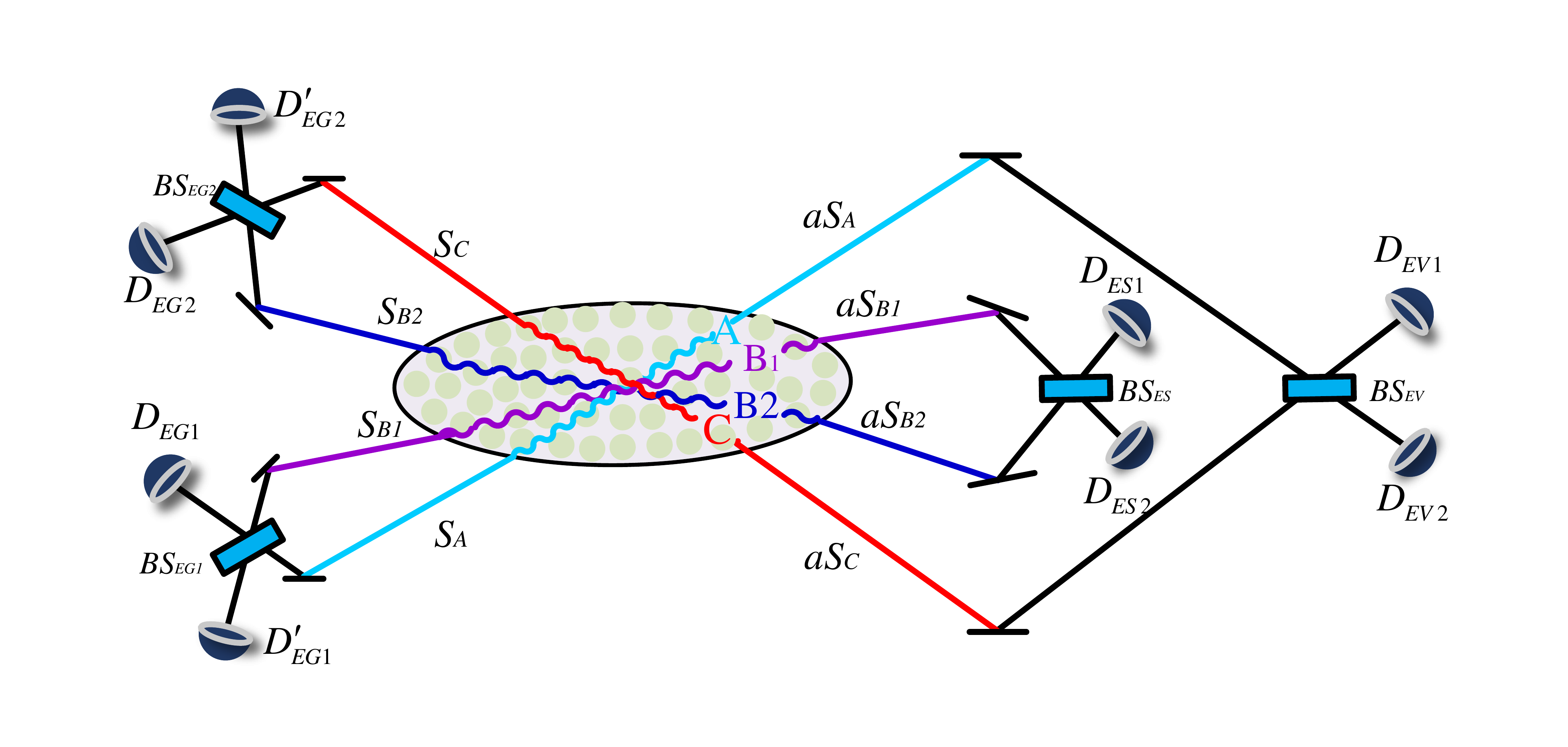}

	\textbf{Fig. 3S.} Schematic view of ES with SW memory ensembles. A, B1, B2, and C represent four memory ensembles. $ S_{A} $, $ S_{B1} $,$  S_{B2} $, and $ S_{C} $ represent Stokes photons. $ aS_{A} $, $ aS_{B1} $, $ aS_{B2} $, and $ aS_{C} $ represent anti-Stokes photons. BSs: 50/50 beam splitters; $ D_{ES1} $, $ D_{ES2} $, … $ D_{EV1} $, $ D_{EV2} $: single-photon detectors.
	\label{S2}
\end{figure}
The experimental setup diagram shown in Fig.1\textit{(a)}, \textit{(d)} and \textit{(e)} in the main text can be simplified into a schematic diagram as shown in Fig. 3S. As pointed in the main text, entanglement states ${\Psi _{A,B1}}(t)$ between A and B1 memory ensembles and ${\Psi _{B2,C}}(t)$  between B2 and C memory ensembles are generated via single-photon detections at the same time in heralded way. Thus, the state of the four memory ensembles is written as: 
\begin{equation}
\begin{array}{c}
		{\Psi _{A,B1,B2,C}} = {\Psi _{A,B1}} \otimes {\Psi _{B2,C}} = \frac{{\rm{1}}}{{\rm{2}}}\left( {\left| {\rm{1}} \right\rangle {}_{\rm{A}}\left| {\rm{0}} \right\rangle {}_{{\rm{B1}}} \pm {{\rm{e}}^{ - i\Delta {\alpha _1}}}\left| {\rm{0}} \right\rangle {}_{\rm{A}}\left| {\rm{1}} \right\rangle {}_{{\rm{B1}}}} \right) \otimes \left( {\left| {\rm{1}} \right\rangle {}_{{\rm{B2}}}\left| {\rm{0}} \right\rangle {}_{\rm{C}} \pm {{\rm{e}}^{ - i\Delta {\alpha _2}}}\left| {\rm{0}} \right\rangle {}_{{\rm{B2}}}\left| {\rm{1}} \right\rangle {}_{\rm{C}}} \right)\\
		= \frac{{\rm{1}}}{{\rm{2}}}(\left| {\rm{1}} \right\rangle {}_{\rm{A}}\left| {\rm{0}} \right\rangle {}_{{\rm{B1}}}\left| {\rm{1}} \right\rangle {}_{{\rm{B2}}}\left| {\rm{0}} \right\rangle {}_{\rm{C}}{\rm{ + }}{{\rm{e}}^{ - i(\Delta {\alpha _1} + \Delta {\alpha _2})}}\left| {\rm{0}} \right\rangle {}_{\rm{A}}\left| {\rm{1}} \right\rangle {}_{{\rm{B1}}}\left| {\rm{0}} \right\rangle {}_{{\rm{B2}}}\left| {\rm{1}} \right\rangle {}_{\rm{C}}\\
		\pm {{\rm{e}}^{ - i\Delta {\alpha _1}}}\left| {\rm{0}} \right\rangle {}_{\rm{A}}\left| {\rm{1}} \right\rangle {}_{{\rm{B1}}}\left| {\rm{1}} \right\rangle {}_{{\rm{B2}}}\left| {\rm{0}} \right\rangle {}_{\rm{C}} \pm {{\rm{e}}^{ - i\Delta {\alpha _2}}}\left| {\rm{1}} \right\rangle {}_{\rm{A}}\left| {\rm{0}} \right\rangle {}_{{\rm{B1}}}\left| {\rm{0}} \right\rangle {}_{{\rm{B2}}}\left| {\rm{1}} \right\rangle {}_{\rm{C}})
	\end{array}
\end{equation}
After a storage time \textit{t1}, we start ES operation between the two links A-B1 and B2-C. For which, we convert the collective excitations in B1 and B2 spin-wave modes into anti-Stokes (read-out) fields $ aS_{B1} $ and $ aS_{B2} $ by applying a read pulse $ R_{1} $. Thus, the state ${\Psi _{A,B1,B2,C}}$ is transferred into the state of two SW modes and two optical fields, which is written as
\begin{equation}
	\begin{array}{l}
		{\Phi _{A,B1,B2,C}} = \frac{{\rm{1}}}{{\rm{2}}}(\left| {\rm{1}} \right\rangle {}_{\rm{A}}\left| {\rm{0}} \right\rangle {}_{{\rm{B1}}}\left| {{{\rm{1}}_{aS}}} \right\rangle {}_{{\rm{B2}}}\left| {\rm{0}} \right\rangle {}_{\rm{C}}{\rm{ + }}{{\rm{e}}^{ - i(\Delta {\alpha _1} + \Delta {\alpha _2})}}\left| {\rm{0}} \right\rangle {}_{\rm{A}}\left| {{{\rm{1}}_{aS}}} \right\rangle {}_{{\rm{B1}}}\left| {\rm{0}} \right\rangle {}_{{\rm{B2}}}\left| {\rm{1}} \right\rangle {}_{\rm{C}}\\
		\begin{array}{*{20}{c}}
			{}&{}&{}&{}&{}&{}&{}&{}
		\end{array} \pm {{\rm{e}}^{ - i\Delta {\alpha _1}}}\left| {\rm{0}} \right\rangle {}_{\rm{A}}\left| {{{\rm{1}}_{aS}}} \right\rangle {}_{{\rm{B1}}}\left| {{{\rm{1}}_{aS}}} \right\rangle {}_{{\rm{B2}}}\left| {\rm{0}} \right\rangle {}_{\rm{C}} \pm {{\rm{e}}^{ - i\Delta {\alpha _2}}}\left| {\rm{1}} \right\rangle {}_{\rm{A}}\left| {\rm{0}} \right\rangle {}_{{\rm{B1}}}\left| {\rm{0}} \right\rangle {}_{{\rm{B2}}}\left| {\rm{1}} \right\rangle {}_{\rm{C}})
	\end{array}
\end{equation}
where, $\left| {{{\rm{1}}_{aS}}} \right\rangle {}_{{\rm{B1}}}$ ($\left| {{{\rm{1}}_{aS}}} \right\rangle {}_{{\rm{B2}}}$ ) represents one photon in the anti-Stokes field $ aS_{B1} $ ($ aS_{B2} $). The $ aS_{B1} $ and $ aS_{B2} $ modes are combined on the beam splitter $ BS_{ES} $. After $ BS_{ES} $, the two output modes $a_{ES1}^\dag  = \left( {a_{a{S_{_{B1}}}}^\dag  + {e^{i\Delta \gamma }}a_{a{S_{_{B2}}}}^\dag } \right)/\sqrt 2 $ and $a_{ES2}^\dag  = \left( {a_{a{S_{_{B1}}}}^\dag  - {e^{i\Delta \gamma }}a_{a{S_{_{B2}}}}^\dag } \right)/\sqrt 2 $  are directed to the detectors $ D_{ES1} $ and $ D_{ES2} $, respectively, where, $a_{ES1}^\dag $  and  $a_{ES2}^\dag $  ($a_{a{S_{_{B1}}}}^\dag $ and $a_{a{S_{_{B2}}}}^\dag $ ) are the creation operators associated with ES1 and ES2 ($ aS_{B1} $ and $ aS_{B2} $) fields, respectively. $\Delta \gamma  = {\gamma _{B1}} - {\gamma _{B2}}$ is the phase difference between the fields $ a{S_{B1}} $ and $ a{S_{B2}} $ from the memory ensembles to $ BS_{ES} $. The state $ {\Phi _{A,B1,B2,C}} $  can be rewritten as:
\begin{equation}
\begin{array}{l}
	{\Phi _{A,B1,B2,C}} = \frac{{\rm{1}}}{{\rm{2}}}\{ \frac{{a_{{\rm{ES1}}}^{\rm{\dag }}\left| {\rm{0}} \right\rangle \left( {\left| {\rm{1}} \right\rangle {}_{\rm{A}}\left| {\rm{0}} \right\rangle {}_{\rm{C}}{\rm{ + }}{{\rm{e}}^{ - i(\Delta {\alpha _1} + \Delta {\alpha _2} + \Delta \gamma )}}\left| {\rm{0}} \right\rangle {}_{\rm{A}}\left| {\rm{1}} \right\rangle {}_{\rm{C}}} \right)}}{{\sqrt {\rm{2}} }}{\rm{ + }}\frac{{a_{{\rm{ES2}}}^{\rm{\dag }}\left| {\rm{0}} \right\rangle \left( {\left| {\rm{0}} \right\rangle {}_{\rm{A}}\left| {\rm{1}} \right\rangle {}_{\rm{C}} - {{\rm{e}}^{ - i(\Delta {\alpha _1} + \Delta {\alpha _2} + \Delta \gamma )}}\left| {\rm{1}} \right\rangle {}_{\rm{A}}\left| {\rm{0}} \right\rangle {}_{\rm{C}}} \right)}}{{\sqrt 2 }}\\
	\begin{array}{*{20}{c}}
		{}&{}&{}
	\end{array} + (\frac{{a_{ES1}^\dag a_{ES1}^\dag \left| {\rm{0}} \right\rangle }}{2}\left| {\rm{0}} \right\rangle {}_{\rm{A}}\left| {\rm{0}} \right\rangle {}_{\rm{C}}{{\rm{e}}^{ - i(\Delta {\alpha _1} + \Delta \gamma )}} - \frac{{a_{ES2}^\dag a_{ES2}^\dag \left| {\rm{0}} \right\rangle }}{2}\left| {\rm{0}} \right\rangle {}_{\rm{A}}\left| {\rm{0}} \right\rangle {}_{\rm{C}}{{\rm{e}}^{ - i(\Delta {\alpha _1} + \Delta \gamma )}})\\
	\begin{array}{*{20}{c}}
		{}&{}&{}
	\end{array} \pm {{\rm{e}}^{i\Delta {\alpha _2}}}\left| {\rm{1}} \right\rangle {}_{\rm{A}}\left| {\rm{0}} \right\rangle {}_{{\rm{B1}}}\left| {\rm{0}} \right\rangle {}_{{\rm{B2}}}\left| {\rm{1}} \right\rangle {}_{\rm{C}}\} 
\end{array}
\end{equation}
Based on the relationships of $a_{{\rm{ES1}}}^{\rm{\dag }}\left| {\rm{0}} \right\rangle {\rm{ = }}\left| {\rm{1}} \right\rangle {}_{{\rm{ES1}}}$  ($a_{{\rm{ES2}}}^{\rm{\dag }}\left| {\rm{0}} \right\rangle {\rm{ = }}\left| {\rm{1}} \right\rangle {}_{{\rm{ES2}}}$) and $ a_{{\rm{ES1}}}^{\rm{\dag }}a_{{\rm{ES1}}}^{\rm{\dag }}\left| {\rm{0}} \right\rangle {\rm{ = }}\sqrt {\rm{2}} \left| {\rm{2}} \right\rangle {}_{{\rm{ES1}}}$ ($a_{{\rm{ES2}}}^{\rm{\dag }}a_{{\rm{ES2}}}^{\rm{\dag }}\left| {\rm{0}} \right\rangle {\rm{ = }}\sqrt {\rm{2}} \left| {\rm{2}} \right\rangle {}_{{\rm{ES2}}}$  )we rewrite the Eq. (S10) as: 
\begin{equation}
\begin{array}{l}
	{\Phi _{A,B1,B2,C}} = \frac{{\rm{1}}}{{\rm{2}}}\{ \frac{{\left| {\rm{1}} \right\rangle {}_{{\rm{ES1}}}\left( {\left| {\rm{1}} \right\rangle {}_{\rm{A}}\left| {\rm{0}} \right\rangle {}_{\rm{C}}{\rm{ + }}{{\rm{e}}^{ - i(\Delta {\alpha _1} + \Delta {\alpha _2} + \Delta \gamma )}}\left| {\rm{0}} \right\rangle {}_{\rm{A}}\left| {\rm{1}} \right\rangle {}_{\rm{C}}} \right)\left| {\rm{0}} \right\rangle {}_{{\rm{ES2}}}}}{{\sqrt {\rm{2}} }}{\rm{ + }}\frac{{\left| {\rm{1}} \right\rangle {}_{{\rm{ES2}}}\left( {\left| {\rm{0}} \right\rangle {}_{\rm{A}}\left| {\rm{1}} \right\rangle {}_{\rm{C}} - {{\rm{e}}^{ - i(\Delta {\alpha _1} + \Delta {\alpha _2} + \Delta \gamma )}}\left| {\rm{1}} \right\rangle {}_{\rm{A}}\left| {\rm{0}} \right\rangle {}_{\rm{C}}} \right)\left| {\rm{0}} \right\rangle {}_{{\rm{ES1}}}}}{{\sqrt 2 }}\\
	\begin{array}{*{20}{c}}
		{}&{}&{}
	\end{array} + {{\rm{e}}^{ - i(\Delta {\alpha _1} + \Delta \gamma )}}\frac{{\left| {\rm{2}} \right\rangle {}_{{\rm{ES1}}}\left| {\rm{0}} \right\rangle {}_{{\rm{ES2}}}}}{{\sqrt 2 }}\left| {\rm{0}} \right\rangle {}_{\rm{A}}\left| {\rm{0}} \right\rangle {}_{\rm{C}} + {{\rm{e}}^{ - i(\Delta {\alpha _1} + \Delta \gamma  + \pi )}}\frac{{\left| {\rm{0}} \right\rangle {}_{{\rm{ES1}}}\left| {\rm{2}} \right\rangle {}_{{\rm{ES2}}}}}{{\sqrt 2 }}\left| {\rm{0}} \right\rangle {}_{\rm{A}}\left| {\rm{0}} \right\rangle {}_{\rm{C}}\\
	\begin{array}{*{20}{c}}
		{}&{}&{}
	\end{array} \pm {{\rm{e}}^{i\Delta {\alpha _2}}}\left| {\rm{1}} \right\rangle {}_{\rm{A}}\left| {\rm{0}} \right\rangle {}_{{\rm{B1}}}\left| {\rm{0}} \right\rangle {}_{{\rm{B2}}}\left| {\rm{1}} \right\rangle {}_{\rm{C}}\} \\
	\begin{array}{*{20}{c}}
		{}&{}&{}
	\end{array}
\end{array}
\end{equation}
where,  $\left| {\rm{1}} \right\rangle {}_{{\rm{ES1}}}$ ($\left| {\rm{1}} \right\rangle {}_{{\rm{ES2}}}$) and $\left| {\rm{2}} \right\rangle {}_{{\rm{ES1}}}$  ($\left| {\rm{2}} \right\rangle {}_{{\rm{ES2}}}$) denotes the state of the field ES1 (ES2) with one photon and two photons, respectively. According to the above Eq.(S11), we see that upon a detection event at the single-photon detector $ D_{ES1} $ or$  D_{ES2} $, the remaining two spin-wave ensembles A and C are projected into the state: 
\begin{equation}
	\rho _{A,C}^{} = \left| 0 \right\rangle \left\langle 0 \right|/2 + \left| {\Psi _{A,C}^{}} \right\rangle \left\langle {\Psi _{A,C}^{}} \right|/2
\end{equation}
where, $\Psi _{A,C}^{} = \left( {\left| 1 \right\rangle {}_A{{\left| 0 \right\rangle }_C} \pm {e^{ - i\xi }}\left| 0 \right\rangle {}_A{{\left| 1 \right\rangle }_C}} \right)/\sqrt 2 $  represents the entangled state between A and C memory ensembles, and  $\left| 0 \right\rangle $ the vacuum part, $\xi  = \Delta {\alpha _1} + \Delta {\alpha _2} + \Delta \gamma $  the sum of the difference phases. In our presented experiment, the pair fields $ S_{A} $ and $ S_{B1} $, the pair fields $ S_{B2} $ and $ S_{C} $, as well as the pair fields $ a{S_{B1}} $ and $ a{S_{B2}} $ have orthogonal polarization, respectively. In the propagations from the memory ensembles to the polarization beam splitters, the fields in each pair go through the same polarization beam displacer and single-mode optical fiber, which are polarization unchanged. 

So,  $\xi $ is passively stabilized and kept constant, which enable us to generate the entangled state $  \Psi _{A,C}^{} $   after ES. Noted that the term $\left| {\rm{1}} \right\rangle {}_{\rm{A}}\left| {\rm{0}} \right\rangle {}_{{\rm{B1}}}\left| {\rm{0}} \right\rangle {}_{{\rm{B2}}}\left| {\rm{1}} \right\rangle {}_{\rm{C}}$ has not contribution to the detection event at $ D_{EV1} $ since we have assumed that the quantum interfaces are ideal case, i.e., ${g_{S,aS}} $ and $ g{'_{S,aS}} \to \infty $. In this way, the vacuum part only influences the success probability, but not the overall fidelity of QR in the ideal case[5]. However, for the case that the quantum interfaces are non-ideal case, for example, ${g_{S,aS}}$ and $g{'_{S,aS}}\sim 40$, the term  will has contribution to the detection event at $ D_{EV1} $, which degrade the entanglement degree and discuses below. 

To demonstrate the entanglement degree of the state $  \Psi _{A,C}^{} $,  we need to measure the visibility $  V _{A,C}^{} $  as well as the suppression parameter $  h _{A,C}^{} $  of the state. For measuring $  V _{A,C}^{} $, we need to convert the collective excitation in the A and C spin-wave modes into the anti-Stokes fields $ aS_{A} $ and $ aS_{C} $, respectively. As mentioned above, such conversions are achieved by applying the read pulse $ R_{2} $ on the atoms at the time t2. In the measurement for the visibility $  V _{A,C}^{} $, we combine the two anti-Stokes fields $ aS_{A} $ and $ aS_{C} $ at a beam splitter $ BS_{EV} $. After $ BS_{EV} $, the two output modes $a_{EV1}^{} = \left( {a_{a{S_A}}^{} + {e^{i\theta }}a_{a{S_{_C}}}^{}} \right)/\sqrt 2 $ and $a_{EV2}^{} = \left( {a_{a{S_{_A}}}^{} - {e^{i\theta }}a_{a{S_{_C}}}^{}} \right)/\sqrt 2 $  are directed to the detectors $ D_{EV1} $ and $ D_{EV2} $, respectively, where, ${a_{a{S_{_A}}}}$ and ${a_{a{S_{_C}}}}$ are the annihilation operators associated with $ aS_{A} $ and $ aS_{C} $ fields, respectively, $\theta$ is the relative phase between the fields $a{S_A}$  and $a{S_C}$  from the memory ensembles to $ BS_{EV} $. Noted that the $ BS_{EV} $ ($ BS_{ES} $) in Fig1 corresponds to $ PBS_{ES} $ ($ PBS_{EV} $) in Fig.2. The visibility $  V _{A,C}^{} $   can be measured as:
\begin{equation}
	{V_{A,C}} = \frac{{{\rm{Max}}\left\{ {{P_{{D_{ES1}},{D_{EV1}}}}(\theta )} \right\} - {\rm{Min}}\left\{ {{P_{{D_{ES1}},{D_{EV1}}}}(\theta )} \right\}}}{{{\rm{Max}}\left\{ {{P_{{D_{ES1}},{D_{EV1}}}}(\theta )} \right\} + {\rm{Min}}\left\{ {{P_{{D_{ES1}},{D_{EV1}}}}(\theta )} \right\}}}
\end{equation}
where, $ {P_{{D_{ES1}},{D_E}_{V1}}} $  denotes the coincidence probability between the detectors $ D_{ES1} $ and $ D_{EV1} $,  ${\rm{Max}}\left\{ {{P_{{D_{ES1}},{D_{EV1}}}}(\theta )} \right\}$ (${\rm{Min}}\left\{ {{P_{{D_{ES1}},{D_{EV1}}}}(\theta )} \right\}$) denotes the maximum (minimum) coincidence when scanning the relative phase $\theta$. 

In the ideal case (i.e., $g_{S,aS}(g'_{S,aS}) \to \infty$) together with a perfect mode overlaps between the retrieved fields from B1 and B2 as well as A and C memory modes, the visibility $V_{A,C}^{} \to 1$. However, in practical experiments, ${g_{S,aS}}$ is imperfect (for example, ${g_{S,aS}}(g'_{S,aS}) \sim 30 - 100$), thus the visibility $  V _{A,C}^{} $  will be less than 1, whose value significantly depended on that of ${g_{S,aS}}$  and $g'_{S,aS}$. Such dependence of $  V _{A,C}^{} $ on ${g_{S,aS}}$   and  $g'_{S,aS}$ is important, which can be used for calculating the threshold of ${g_{S,aS}}$   and  $g'_{S,aS}$  for achieving entanglement  (${{\cal C}_{A,C}} > 0$) between A and C spin waves. As described in Eq.S9, the state of the four memory modes $ {\Psi _{A,B1,B2,C}} $  involves in the four terms, each of them correspond to the different storage cases of the spin-wave excitations in A, B1, B2, and C modes. We now calculate the $ {P_{{D_{ES1}},{D_{EV1}}}}(\theta ) $  by considering the following three cases that include the contributions of the four terms.

\textbf{(1)	The contribution of the first case to the coincidence probability } 

In the first case, we consider the contributions of the terms  $\left| {\rm{1}} \right\rangle {}_{\rm{A}}\left| {\rm{0}} \right\rangle {}_{{\rm{B1}}}\left| {\rm{1}} \right\rangle {}_{{\rm{B2}}}\left| {\rm{0}} \right\rangle {}_{\rm{C}}{\rm{ + }}\left| {\rm{0}} \right\rangle {}_{\rm{A}}\left| {\rm{1}} \right\rangle {}_{{\rm{B1}}}\left| {\rm{0}} \right\rangle {}_{{\rm{B2}}}\left| {\rm{1}} \right\rangle {}_{\rm{C}}$ to the  $ {P_{{D_{ES1}},{D_{EV1}}}}(\theta ) $, where, the first term corresponds to that only spin waves A and B2 store 1 excitation, and B1 and C store 0 excitation, respectively; the second term refers to that the modes B1, C store 1 excitation, and A and B2 store 0 excitation, respectively. The above terms can be rewritten as the following superposition $\frac{{\left| {\rm{1}} \right\rangle {}_{{\rm{ES1}}}\left( {\left| {\rm{1}} \right\rangle {}_{\rm{A}}\left| {\rm{0}} \right\rangle {}_{\rm{C}}{\rm{ + }}{{\rm{e}}^{ - i\xi }}\left| {\rm{0}} \right\rangle {}_{\rm{A}}\left| {\rm{1}} \right\rangle {}_{\rm{C}}} \right)\left| {\rm{0}} \right\rangle {}_{{\rm{ES2}}}}}{{\sqrt {\rm{2}} }}{\rm{ + }}\frac{{\left| {\rm{1}} \right\rangle {}_{{\rm{ES2}}}\left( {\left| {\rm{0}} \right\rangle {}_{\rm{A}}\left| {\rm{1}} \right\rangle {}_{\rm{C}} - {{\rm{e}}^{ - i\xi }}\left| {\rm{1}} \right\rangle {}_{\rm{A}}\left| {\rm{0}} \right\rangle {}_{\rm{C}}} \right)\left| {\rm{0}} \right\rangle {}_{{\rm{ES2}}}}}{{\sqrt 2 }}$. This superposition shows that the detection of a photon at DES1 will project the two spin-wave modes A and C into the entangled state $\Psi _{A,C}^{( + )} = \left( {\left| 1 \right\rangle {}_A{{\left| 0 \right\rangle }_C} + \left| 0 \right\rangle {}_A{{\left| 1 \right\rangle }_C}} \right)/\sqrt 2 $.  The probability of detecting such a photon at $ D_{ES1} $ is $P_{{D_{ES1}}}^{} = \eta \gamma ({t_1})$, which represents the probability to herald a successful ES for the ideal case. For verifying entanglement between A and C spin-wave memory, we map the spin waves stored A and C nodes into the anti-Stokes fields $ aS_{A} $ and $ aS_{C} $, respectively. The spin-wave entangled state $ {\Psi _{A,C}} $   are then transformed into a single-photon entangled state ${\Phi _{A,C}}$  between $ aS_{A} $ and $ aS_{C} $ modes, which is write as  
\begin{equation}
	\Phi _{A,C}^{} \propto \sqrt \gamma  \left( {{a_{a{S_A}}}^\dag \left| 0 \right\rangle  \pm {e^{ - i\xi }}{a_{a{S_C}}}^\dag \left| 0 \right\rangle } \right)/\sqrt 2 
\end{equation}
$  {{a_{a{S_A}}}^\dag} $ ($  {{a_{a{S_C}}}^\dag} $) refers to the creation operator associated with anti-Stokes fields $ aS_{A } $($ aS_{C} $). We combine the modes aSA and aSC on the beam splitter $ BS_{EV} $. After $ BS_{EV} $, the anti-Stokes output modes $ {a_{EV1}} = \left( {{a_{a{S_A}}} + {e^{i\theta }}{a_{a{S_C}}}} \right)/\sqrt 2 $   and ${a_{EV2}} = \left( {{a_{a{S_A}}} - {e^{i\theta }}{a_{a{S_C}}}} \right)/\sqrt 2 $  are directed to the detectors $ D_{EV1} $ and $ D_{EV2} $, respectively. As mentioned above, $\theta$  is the phase difference between the two modes $a{S_A}$   and  $a{S_C}$ , which can be adjusted by a phase adjustor (PA). We rewrite the single-photon entanglement state as: 
\begin{equation}
	{\Phi _{A,C}} = \frac{{\sqrt {\gamma (t)} }}{2}\left( {a_{EV1}^\dag (1 + {e^{ - i(\xi  + \theta )}}) + a_{EV2}^\dag (1 - {e^{ - i(\xi  + \theta )}})} \right)\left| 0 \right\rangle 
\end{equation}
In the measurement of the single-photon entangled state, the sum of phase difference $\xi $  is kept unchanged, while the relative phase  $ \theta $ is scanned. The probability of detecting a photon at the detector $ D_{EV1} $ may be expressed as[1]: 
\begin{equation}
	P_{{D_{EV1}}}^{}(\theta ,{t_2}) = \eta \gamma ({t_2})\left\langle {{{\left| {1 + {e^{ - i\theta }}} \right|}^2}} \right\rangle /4 = \eta \gamma ({t_2})\left( {1 + \cos \theta } \right)/2 
\end{equation}
where, we have made a translation of $\xi  + \theta  \to \theta $ ,  $\eta $ is the efficiency for either detection channel, with ${\eta ^{(A)}} = {\eta ^{(B1)}} = {\eta ^{(B2)}} = {\eta ^{(C)}} = \eta $. In our practical ES experiment, however, each of the quantum interfaces is the non-ideal, i.e.,  there are the unconditional detection events described by $P_{aS}^{(X)} = \frac{{\gamma (t)\eta }}{{g_{S,aS}^{(X)} - 1}} \approx \frac{{\gamma (t)\eta }}{{g_{S,aS}^{(X)}}}$  (see Eq.S5) in the $ aS_{X} $ (X=A, B1, B2, C) in the read-out channels. Such unconditional detection events result from noise photons in $ aS_{X} $ channels (modes). The noise photons in the $ aS_{A} $ and $ aS_{C} $ modes also direct into the EV1 and EV2 field modes and then induce an unconditional detection probability at $ D_{EV1} $, which is written as 
\begin{equation}
	P_{{D_{EV1}}}^{}({t_2}) = \frac{1}{2} \times \left[ {P_{aS}^{(A)}({t_2}) + P_{aS}^{(C)}({t_2})} \right] = {P_{aS}}'({t_2})
\end{equation}
where, the factor 1/2 correspond to the 50\% chance that the noise photon is reflected or transmitted at the beam splitter ($ BS_{V} $). Considering the above two components described at [S16] and [S17], we write the coincidence probability between the read-out fields in ES1 and EV1 modes as[1] 
\begin{equation}
	P_{{D_{ES1,}}{D_{_{EV1}}}}^{} = P_{{D_{ES1}}}^{}({t_1})P_{{D_{EV1}}}^{}(\theta ,{t_2}) + P_{{D_{ES1}}}^{}({t_1})P_{{D_{EV1}}}^{}({t_2}) = \gamma \left( {{t_1}} \right)\eta  * \eta \gamma ({t_2})\left( {1 + \cos \theta } \right)/2 + \gamma \left( {{t_1}} \right)\eta {P_{aS}}'({t_2})
\end{equation}
which represents a contribution of the terms ($\left| {\rm{1}} \right\rangle {}_{\rm{A}}\left| {\rm{0}} \right\rangle {}_{{\rm{B1}}}\left| {\rm{1}} \right\rangle {}_{{\rm{B2}}}\left| {\rm{0}} \right\rangle {}_{\rm{C}}{\rm{ + }}\left| {\rm{0}} \right\rangle {}_{\rm{A}}\left| {\rm{1}} \right\rangle {}_{{\rm{B1}}}\left| {\rm{0}} \right\rangle {}_{{\rm{B2}}}\left| {\rm{1}} \right\rangle {}_{\rm{C}}$) to the coincidence probability between the detectors $ D_{ES1} $ and $ D_{EV1} $.

On the other hand, there are the unconditional detection probabilities $P_{aS}^{} = \frac{{\gamma (t)\eta }}{{g_{S,aS}^{} - 1}} \approx \frac{{\gamma (t)\eta }}{{g_{S,aS}^{}}}$  (see Eq.S5) in the $ aS_{B1} $ and $ aS_{B2} $ channels during the retrieval in ES process, which corresponds to the probability detecting noise photons in $ aS_{B1} $ and $ aS_{B2} $ channels (modes). The noise photons in $ aS_{B1} $ and $ aS_{B2} $ channels will direct into the ES1 channel via the beam splitter $ B_{ES} $, leads to an unconditional detection probability  $P_{{D_{ES1}}}^{{\rm{noise}}}({t_1}) = \frac{1}{2} \times \left[ {P_{aS}^{}({t_1}) + P_{aS}^{}({t_1})} \right] = {P_{aS}}({t_1})$  at $ D_{ES1} $, where, the factor 1/2 results from the beam-splitter ratio of 50-50\%. If such an unconditional detection event is occurred at $ D_{ES1} $, the A and C spin waves are projected in a separable state instead of the entangled state $ {\Psi _{A,C}} $. In this case, the A and C spin waves have no coherence. For the incoherent retrievals, the detection probability at $ D_{EV1} $ can be calculated according to the terms $\left| {\rm{1}} \right\rangle {}_{\rm{A}}\left| {\rm{0}} \right\rangle {}_{\rm{C}}{\rm{ + }}\left| {\rm{0}} \right\rangle {}_{\rm{A}}\left| {\rm{1}} \right\rangle {}_{\rm{C}}$, which is $P_{{D_{EV1}}}^{(inc - R)}({t_2}) = \frac{1}{2} \times \left[ {2 \times \eta \gamma ({t_2})} \right] = \eta \gamma ({t_2})$, where, the factor 1/2 correspond to the 50\% chance that the read-out photons from A and C is reflected or transmitted at the beam splitter ($ BS_{V} $), while the factor 2 result from the symmetry of the scheme where the photon can come from either memory. These detections at $ D_{ES1} $ and $ D_{EV1} $ contribute an accidental coincidence to the coincident probability $P_{{D_{ES1}},{D_{EV1}}}^{}$, which is written as 
\begin{equation}
	P_{{D_{ES1}}}^{noise}({t_2})P_{{D_{_{EV1}}}}^{(inc - R)}({t_2}) = {P_{aS}}({t_1})\eta \gamma ({t_2})
\end{equation}
Considering the two contributions in equations (S18) and (S19), the coincidence probability $P_{{D_{ES1}},{D_{EV1}}}^{}(\theta ,{t_2})$  for the first case may be written as:
\begin{equation}
	\begin{array}{l}
		P_{{D_{ES1}},{D_{EV1}}}^{(1 - {\rm{th}})}(\theta ,{t_2}) = {P_{ES1}}({t_1})P(\theta ,{t_2}) + P_{ES1}^{}({t_1})P_{{D_{V1}}}^{noise}({t_2}) + P_{{D_{ES1}}}^{noise}({t_1})P_{{D_{EV1}}}^{(inc - R)}({t_2})\\
		\begin{array}{*{20}{c}}
			{}&{}&{}&{}&{}&{}&{}&{}&{}&{}&{}&{}
		\end{array} = \gamma \left( {{t_1}} \right)\eta \left( {\eta \gamma ({t_2})\left( {1 + \cos \theta } \right)/2} \right) + {P_{aS}}({t_1})\eta \gamma \left( {{t_2}} \right) + \eta \gamma \left( {{t_1}} \right)P{'_{aS}}({t_2})
	\end{array}
\end{equation}
\textbf{(2)	 The contribution of the second case to the coincidence probability  ${P_{{D_{ES1}},{D_E}_{V1}}}$ }

The second case refers to the one where B1 and B2 (A and C) memories store 1 (0) excitation, respectively. This storage case corresponds to the term  $\left| {\rm{0}} \right\rangle {}_{\rm{A}}\left| {\rm{1}} \right\rangle {}_{{\rm{B1}}}\left| {\rm{1}} \right\rangle {}_{{\rm{B2}}}\left| {\rm{0}} \right\rangle {}_{\rm{C}}$ in Eq.S8, which results in the vacuum part in Eq.(S12). For this case, the probability of detecting a photon at $ D_{EV1} $ is $P_{ES1}^{2 - {\rm{th}}}({t_1}) \approx \gamma ({t_1})\eta $  and the probability of detecting a photon at $ D_{EV1} $ is ${P_{{D_{EV1}}}}({t_2}) = {P_{aS}}'({t_2})$. The two detections give rise an accident coincidence contributing to the coincidence probability between $ D_{EV1} $ and $ D_{EV1} $, which is
\begin{equation}
	P_{{D_{ES1}},{D_{EV1}}}^{(2 - th)} = \gamma \left( {{t_1}} \right)\eta  * {P_{aS}}'({t_2})
\end{equation}
\textbf{(3)	 The contribution of the third case to the coincidence probability ${P_{{D_{ES1}},{D_E}_{V1}}}$  }

The third case refers to the one in which B1 and B2 nodes store 0 excitation, while A and C nodes store 1 excitation, respectively. This case correspond the term $\left| {\rm{1}} \right\rangle {}_{\rm{A}}\left| {\rm{0}} \right\rangle {}_{{\rm{B1}}}\left| {\rm{0}} \right\rangle {}_{{\rm{B2}}}\left| {\rm{1}} \right\rangle {}_{\rm{C}}$. For the case that B1 and B2 quantum interfaces are in ideal case, i.e.,$\frac{{{P_{aS}}({t_1})}}{{\eta \gamma }} \sim 0$, this case is excluded the state ${\rho _{A,C}}$ after ES because the probability of detecting a photon at $ D_{ES1} $ can be neglected. However, in the practical experiments, these QIs are non-ideal, and then the noise photons in the retrieved fields can’t be neglected. Thus, at $ D_{ES1} $, one will detect unconditional photon, the detection probability is $P_{{D_{ES1}}}^{3 - {\rm{th}}}({t_1}) = \frac{1}{2} \times \left[ {P_{aS}^{(B1)}({t_1}) + P_{aS}^{(B2)}({t_1})} \right] = {P_{aS}}({t_1})$. While, at $ D_{EV1} $, the probability of detecting a retrieved photon is ${P_{{D_{EV1}}}}({t_2}) \approx \gamma \left( {{t_2}} \right)\eta $. The two detections give rise an accident coincidence, which contribute to the coincidence probability between $ D_{EV1} $ and $ D_{EV1} $ and can be 
\begin{equation}
P_{{D_{ES1}},{D_{EV1}}}^{(3 - th)}(\theta ,{t_2}) \simeq {P_{aS}}({t_1}) * \gamma \eta ({t_2})
\end{equation}
Finally, considering the contributions of the $ P_{{D_{ES1}},{D_{EV1}}}^{i - th} $  in three case, we write the total coincidence probability as
\begin{equation}
	P_{{D_{ES1}},{D_{EV1}}}^{}(\theta ,{t_2}) = \sum\limits_{i = 1}^3 {P_{{D_{ES1}},{D_{EV1}}}^{(i - th)}(\theta ,{t_2})}  = \eta \gamma ({t_1})*\eta \gamma ({t_2})\frac{{(1 + \cos \theta )}}{2} + 2\eta \gamma ({t_2})*{P_{aS}}({t_1}) + 2\eta \gamma ({t_1})*{P_{aS}}'({t_2})
\end{equation}
So, the visibility $ {V_{A,C}} $ is expressed as
\begin{equation}
{V_{A,C}} = \frac{{{P_{{D_{ES1}},{D_{EV1}}}}(0) - {P_{{D_{ES1}},{D_{EV1}}}}(\pi )}}{{{P_{{D_{ES1}},{D_{EV1}}}}(0) + {P_{{D_{ES1}},{D_{EV1}}}}(\pi )}} = \frac{1}{{1 + \frac{4}{{g{'_{S,aS}}({t_2}) - 1}} + \frac{4}{{{g_{S,aS}}({t_1}) - 1}}}} \approx 1 - \frac{4}{{g{'_{S,aS}}({t_2})}} - \frac{4}{{{g_{S,aS}}({t_1})}}	
\end{equation}

\section{IV.	Dependence of suppression parameter h of the entangled state $  {\Psi _{A,C}}(t) $  on the cross-correlation functions of the quantum interfaces}
As motioned in the main text, the suppression parameter of the entangled state $  {\Psi _{A,C}}(t) $   is written as ${h_{A,C}} = \frac{{{p_{11}}}}{{{p_{10}}{p_{01}}}}$ with $ {p_{ij}} $  denoting the probability of detecting $i \in \left( {0,1} \right)$  photons in the retrieved field $ aS_{A} $, and  $j \in \left( {0,1} \right)$  photons in the retrieved field $ aS_{C} $. In the following, we calculate the contributions of the four terms in Eq.S8, to the probabilities of ${p_{10}}$, ${p_{01}}$, and  ${p_{11}}$  respectively. Such calculations are divided into the following four cases.

(1)	 The first case corresponds to the contribution of the term $\left| {\rm{1}} \right\rangle {}_{\rm{A}}\left| {\rm{0}} \right\rangle {}_{{\rm{B1}}}\left| {\rm{1}} \right\rangle {}_{{\rm{B2}}}\left| {\rm{0}} \right\rangle {}_{\rm{C}}$ , where, B2 and A (B2 and C) memories store 1 (0) collective excitation, respectively. In the ES operation, the collective excitation in B1 memory is retrieved, and the probability of detecting the retrieved photon at $ D_{ES1} $ is $P_{ES1}^{1 - {\rm{th}}}({t_1}) = \gamma \left( {{t_1}} \right)\eta /2$. In the measurement for h, we converted the 1 collective excitation in A memory into $ aS_{A} $ mode and 0 excitation in C memory into $ aS_{C} $ mode. So, the contribution of this case to the three probabilities are $p_{10}^{(1 - {\rm{st}})} = \eta \gamma ({t_2})$,  $p_{01}^{(1 - {\rm{st}})} \approx 0$, and $p_{11}^{(1 - {\rm{st}})} = \eta \gamma ({t_2})P{'_{aS}}({t_2})$.

(2)	The second case corresponds to the term $\left| {\rm{0}} \right\rangle {}_{\rm{A}}\left| {\rm{1}} \right\rangle {}_{{\rm{B1}}}\left| {\rm{0}} \right\rangle {}_{{\rm{B2}}}\left| {\rm{1}} \right\rangle {}_{\rm{C}}$, where, B1 and C (B2 and A) memories store 1 (0) collective excitation, respectively. In the ES operation, the collective excitation in B2 memory is retrieved. The probability of detecting the retrieved photon at $ D_{ES1} $ is $P_{ES1}^{2 - {\rm{nd}}}({t_1}) = \gamma \left( {{t_1}} \right)\eta /2$. In the measurement for h, we converted the 0 (1) collective excitation in A (C) memory into $ aS_{A} $ ($ aS_{C} $) mode. So, the contribution of this case to the three probabilities are $p_{10}^{(2 - {\rm{nd}})} = 0$, $p_{01}^{(2 - {\rm{nd}})} = \eta \gamma ({t_2})$, $p_{11}^{(2 - {\rm{nd}})} = \eta \gamma ({t_2}) * P{'_{aS}}({t_2})$. 

(3)	The third case corresponds to the term $ \left| {\rm{0}} \right\rangle {}_{\rm{A}}\left| {\rm{1}} \right\rangle {}_{{\rm{B1}}}\left| {\rm{1}} \right\rangle {}_{{\rm{B2}}}\left| {\rm{0}} \right\rangle {}_{\rm{C}}$ , where, B1 and B2 (C and A) memories store 1 (0) collective excitation, respectively. In the ES operation, the collective excitation in B2 memory is retrieved and the probability of detecting the retrieved photon at $ D_{ES1} $ is $P_{ES1}^{3 - {\rm{rd}}}({t_1}) \approx \gamma \left( {{t_1}} \right)\eta $ . The contributions of the this case to the three probabilities are $p_{01}^{(3 - {\rm{rd}})} = p_{10}^{(3 - {\rm{rd}})} \approx 0$, $p_{11}^{(3 - {\rm{rd}})} \approx 0$.

(4)	The fourth case corresponds to the term $\left| {\rm{1}} \right\rangle {}_{\rm{A}}\left| {\rm{0}} \right\rangle {}_{{\rm{B1}}}\left| {\rm{0}} \right\rangle {}_{{\rm{B2}}}\left| {\rm{1}} \right\rangle {}_{\rm{C}}$, where, C and A (B1 and B2) memories store 1 (0) collective excitation, respectively. In the ES operation, there is 0 collective excitation in B1 and B2 memories to be retrieved and the probability of detecting a photon at $ D_{ES1} $ is only involved in the noise photon, which is $P_{ES1}^{4 - {\rm{th}}}({t_1}) \approx {P_{aS}}({t_1})$ . The contributions of the this case to the three probabilities are $p_{01}^{(4 - {\rm{th}})} = p_{10}^{(4 - {\rm{th}})} = \eta \gamma ({t_2})[1 - \eta \gamma ({t_2})] \approx \eta \gamma ({t_2})$, $p_{11}^{(4 - {\rm{th}})} = {[\eta \gamma ({t_2})]^2}$. 
The normalized probability of occurring the four cases correspond to the normalized probability of detecting a photon at $ D_{ES1} $, which are 
$P_{Nor}^{^{(1 - {\rm{st}})}} = P_{ES1}^{^{(1 - {\rm{st}})}}/\sum\limits_{i = 1}^4 {P_{ES1}^{^{(i - {\rm{th}})}}}  \buildrel\textstyle.\over= 1/4$;  $P_{Nor}^{^{(2 - {\rm{nd}})}} = P_{ES1}^{^{(2 - {\rm{nd}})}}/\sum\limits_{i = 1}^4 {P_{ES1}^{^{(i - {\rm{th}})}}}  \buildrel\textstyle.\over= 1/4$;  $P_{Nor}^{^{(3 - {\rm{rd}})}} = P_{ES1}^{^{(3 - {\rm{rd}})}}/\sum\limits_{i = 1}^4 {P_{ES1}^{^{(i - {\rm{th}})}}}  \buildrel\textstyle.\over= 2/4$; $P_{Nor}^{^{(4 - {\rm{th}})}} = P_{ES1}^{^{(4 - {\rm{th}})}}/\sum\limits_{i = 1}^4 {P_{ES1}^{^{(i - {\rm{th}})}}}  \buildrel\textstyle.\over= 1/(2{g_{S,aS}}({t_1})$, 
respectively. The total probability of detecting one photon in the retrieved field $ aS_{A} $ ($ aS_{C} $), and 0 photon in the retrieved field $ aS_{C} $ ($ aS_{A} $) is ${p_{10}} = \sum\limits_{i = 1}^4 {P_{Nor}^{(i - {\rm{th}})}p_{10}^{(i - {\rm{th}})}}  \approx \frac{{\eta \gamma ({t_2})}}{4}$(${p_{01}} = \sum\limits_{i = 1}^4 {\left( {P_{Nor}^{(i - {\rm{th)}}}p_{01}^{(i - {\rm{th}})}} \right)}  \approx \frac{{\eta \gamma ({t_2})}}{4}$), the probability of detecting two-photon coincidence between in the retrieved fields $ aS_{A} $ and $ aS_{C} $ is ${p_{11}} = \sum\limits_i^4 {\left( {P_{Nol}^{(i - {\rm{th}})}p_{11}^{(i - {\rm{th}})}} \right)}  = {[\eta \gamma ({t_2})]^2}\left[ {1/\left( {2{g_{S,aS}}({t_1})} \right) + 1/\left( {2g{'_{S,aS}}({t_2})} \right)} \right]$. 

In this way, the suppression h of the entangled state  ${\Psi _{A,C}}(t)$ can be evaluated as  
\begin{equation}
{h_{A,C}} = \frac{{{p_{11}}}}{{{p_{10}}{p_{01}}}} \approx 8/\left( {\frac{1}{{{g_{S,aS}}({t_1})}} + \frac{1}{{g{'_{S,aS}}({t_2})}}} \right)
\end{equation}

[1] M. Wang, H. Jiao, J. Lu, W. Fan, Z. Yang, M. Xi, S. Li, H. Wang, Cavity-enhanced and spatial-multimode spin-wave-photon quantum interface. arXiv:2307.12523.

[2] C. Sun, Y. Li, Y. Hou, M. Wang, S. Li, H. Wang, Decoherence of Single‐Excitation Entanglement over Duan‐Lukin‐Cirac‐Zoller Quantum Networks Caused by Slow‐Magnetic‐Field Fluctuations and Protection Approach. Advanced Quantum Technologies, 2300148 (2023).

[3] L. Heller, P. Farrera, G. Heinze, H. de Riedmatten, Cold-Atom Temporally Multiplexed Quantum Memory with Cavity-Enhanced Noise Suppression. Phys. Rev. Lett. 124, 210504 (2020).

[4] S. Chen, Y. A. Chen, T. Strassel, Z. S. Yuan, B. Zhao, J. Schmiedmayer, J. W. Pan, Deterministic and storable single-photon source based on a quantum memory. Phys. Rev. Lett. 97, 173004 (2006).

[5] J. Laurat, C.-W. Chou, H. Deng, K. S. Choi, D. Felinto, H. de Riedmatten, H. J. Kimble, Towards experimental entanglement connection with atomic ensembles in the single excitation regime. New J. Phys. 9, 207 (2007).

\end{widetext}
\end{document}